\documentclass[aps,prb,preprint,groupedaddress,showpacs]{revtex4}

\usepackage{graphicx}
\usepackage{dcolumn}
\usepackage{bm}
\usepackage{subfigure}

\begin{document}


\title{Finite-Size Scaling Critical Behavior of Randomly Pinned Spin-Density Waves}


\author{Ronald Fisch}
\email[]{ron@princeton.edu}
\affiliation{382 Willowbrook Dr.\\
North Brunswick, NJ 08902}


\date{\today}

\begin{abstract}
We have performed Monte Carlo studies of the 3D $XY$ model with random
uniaxial anisotropy, which is a model for randomly pinned spin-density
waves.  We study $L \times L \times L$ simple cubic lattices, using $L$
values in the range 16 to 64, and with random anisotropy strengths of
$D / 2 J$ = 1, 2, 3, 6 and $\infty$.  There is a well-defined finite
temperature critical point, $T_c$, for each these values of $D / 2 J$.
We present results for the angle-averaged magnetic structure factor,
$S ( k )$ at $T_c$ for $L = 64$.  We also use finite-size scaling analysis
to study scaling functions for the critical behavior of the specific heat,
the magnetization and the longitudinal magnetic susceptibility.  Good
data collapse of the scaling functions over a wide range of $T$ is seen
for $D / 2 J$ = 6 and $\infty$.  For our finite values of $D / 2 J$ the
scaled magnetization function increases with $L$ below $T_c$, and appears
to approach an $L$-independent limit for large $L$.  This suggests that
the system is ferromagnetic below $T_c$.

\end{abstract}


\maketitle

\section{Introduction}

The Harris-Plischke-Zuckermann model\cite{HPZ73} has long been used to study the
effects of random uniaxial anisotropy on ferromagnetism.  The Hamiltonian of this
random anisotropy model (RAM) is
\begin{equation}
  H_{RAM} ~=~ - J \sum_{\langle ij \rangle} \vec{\bf S}_{i} \cdot \vec{\bf S}_{j}
  ~-~ D \sum_{i} ( \hat{\bf n}_{i} \cdot \vec{\bf S}_{i} )^2   \, ,
\end{equation}
where each $\vec{\bf S}_{i}$, the dynamical on site $i$, is usually taken to be
a classical three-component spin of unit length.  Each $\hat{\bf n}_{i}$ is a
time-independent unit vector.  The $\hat{\bf n}$ on different sites are assumed
to be uncorrelated random variables.  $\sum_{\langle ij \rangle}$ is a sum over
nearest neighbors on some lattice.  In this work we will use a simple cubic
lattice with periodic boundary conditions, and we will study the case of
two-component ($n = 2$) spins.

As was discussed in some detail in an earlier paper,\cite{Fis95} if one chooses
the $\vec{\bf S}_{i}$ and the $\hat{\bf n}_{i}$ to be two-component vectors,
then the Hamiltonian can be mapped onto a model of a spin-density wave (SDW) in
an anisotropic material with an easy axis.  For $XY$ spins, {\it i.e.} $n = 2$,
the Hamiltonian of the model may be rewritten as
\begin{equation}
  H ~=~ - J \sum_{\langle ij \rangle} \cos ( \phi_{i} - \phi_{j} )
  ~-~ {D \over 2} \sum_{i} [ \cos ( 2 ( \phi_{i} - \theta_{i} ) )
  ~-~ 1 ]  \, .
\end{equation}
Each $\phi_{i}$ is a dynamical variable which takes on values between 0
and $2 \pi$. The $\langle ij \rangle$ indicates here a sum over nearest
neighbors on a simple cubic lattice of size $L \times L \times L$.  We
choose each $\theta_{i}$ to be an independent identically distributed
quenched random variable, with the probability distribution
\begin{equation}
  P ( \theta_i ) ~=~ 1 / 2 \pi   \,
\end{equation}
for $\theta_i$ between 0 and $2 \pi$.  A constant term has also been
added to the anisotropy, to make the Hamiltonian well-behaved in the
limit $ D / J \to \infty $.

In this work we will study Eqn.~2 on the simple cubic lattice over a range
of $D / J$, using Monte Carlo simulations.  The large increase in available
computing resources over the last fifteen years makes possible significant
improvements over the earlier results.\cite{Fis95}  By studying a range of
$L$, we will be able to learn about the stability of long-range order against
random pinning which respects the Kramers degeneracy, such as alloy disorder,
and the critical behavior of a SDW in an easy-axis material with this type of
pinning.

\section{Random pinning effects}

In the limit $ D / J \to \infty $, often called the Ising limit, both
analytical\cite{DV80,HCB87,FH90} and numerical\cite{JK80,Fis95,TPV06,LLMPV07}
calculations become substantially simplified.  This is due to the fact that
in this Ising limit the random anisotropy term in the Hamiltonian becomes
a projection operator, and each spin has only two allowed states.  It has
been argued that for large $D / J$ the behavior is close to the
$D / J = \infty$ limit as long as $T \ll D$.\cite{HCB87}  It has also been
found, however, that for $n = 2$ at low temperatures and moderately
large values of $D / J$ the magnetization per spin on $L \times L \times L$
simple cubic lattices, $|\vec{M} ( L )|$, decreases\cite{Fis95} as the
temperature, $T$, is lowered.  This effect was not seen for $D / J = \infty$.

A similar effect is seen in the case of the random bond Ising model (RBIM),
where the Nishimori gauge symmetry causes the magnetization to have a maximum
at a finite $T$ on the Nishimori line.\cite{Nis81,HTPV07}  The RBIM is the
natural extension\cite{CL77,Fis90} of the RAM to the case of Ising spins,
$n = 1$.  Thus it should be expected that the phase diagram of the $n = 2$
RAM has a close relation to that of the RBIM.  However, there are aspects of
the phase diagrams which remain somewhat mysterious.  For example, Chen and
Lubensky\cite{CL77} found that the critical exponents which describe the
stability of the ferromagnet-spin glass-paramagnet multicritical point for
the random bond model in $6 - \epsilon$ dimensions are well-behaved for
$n = 1$, but become complex for $n = 2$ and $n = 3$.  One interpretation of
this puzzling result is that the multicritical point itself becomes unstable
in $6 - \epsilon$ dimensions, so that it becomes a region of the phase
diagram, rather than a single point.  In this expanded multicritical region
one might expect to find quasi-long-range order (QLRO).  Although an explicit
calculation has not been done, a similar result would not be surprising
for the RAM.  The existence of QLRO in the RAM was first suggested by
Aharony and Pytte\cite{AP80} in 1980.  They later\cite{AP83} pointed out that
higher order terms might make the correlation length, $\xi$, finite below
$T_c$.  Feldman\cite{Fel01} has argued that QLRO should be common in
disordered magnets and similar systems.

Thus there are a number of possibilities available for the topology of the
phase diagram.  In a Cayley-tree mean-field theory, where QLRO does not occur,
it is known\cite{HCB87} that in the limit $D / J \to \infty$ the phase
diagram depends on the parameter $z / n$, where $z$ is the number of nearest
neighbor spins.  Thus it is to be expected that the phase diagram in three
dimensions will also depend on the lattice type, $n$ and the range of the
exchange interactions, as well as on $T / J$ and $D / J$.  For the simple
cubic lattice (which has $z = 6$) it has been shown\cite{LLMPV07} that in the
limit $D / J \to \infty$ the ground state is an Ising spin glass when
$n \ge 3$.  For small $D / J$, however, where one does not expect the
qualitative behavior to depend on $z$, Feldman\cite{Fel01} predicts QLRO in
$d = 3$ for $n \le 4$.  In the $n = 3$ case, this appears to be confirmed by
Monte Carlo calculations.\cite{Fis98}

The presence of a reentrant phase is difficult to demonstrate conclusively
using the type of numerical calculations we have performed here.  It was only
relatively recently that reentrance was demonstrated convincingly in the
$d = 2$ RBIM.\cite{WHP03,HTPV08}  There may also be a range of $D / J$ for
which the three-dimensional $n = 2$ model has a reentrant ferromagnetic
phase.  One motivation for believing this is that reentrance is frequently
observed in laboratory experiments.  Another is the work of Pelcovits, Pytte
and Rudnick\cite{PPR78,Pel79} who argue that ferromagnetism should be unstable
in the RAM for low $T$ and small $D / J$.  Since magnetization can increase
with increasing $T$ at low $T$, (which was not known at the time of their
work,) it is not correct to claim that the absence of ferromagnetism near
$T = 0$ precludes the existence of a ferromagnetic phase in the RAM at a
somewhat higher $T$.

Larkin\cite{Lar70} studied a model for a vortex lattice in a type-II
superconductor.  His model replaces the spin-exchange term of the
Hamiltonian with a harmonic potential, so that each $\phi_{i}$ is no
longer restricted to lie in a compact interval.  He argued that for any
non-zero value of a random field this model has no long-range order on a
lattice whose dimension $d$ is less than or equal to four.  This argument,
using the harmonic potential instead of the spin-exchange, is only
rigorously correct in the limit $n \to \infty$.

A more intuitive derivation of the result was given by Imry and
Ma,\cite{IM75} who assumed that the increase in the energy of an $L^d$
lattice when the order parameter is twisted at a boundary scales as
$L^{d - 2}$, just as it does in the nonrandom ferromagnet.  As argued
by Imry and Ma,\cite{IM75} and later justified more
carefully,\cite{AIM76,PS82} within an $\epsilon$-expansion one finds
the phenomenon of ``dimensional reduction". Within this perturbation
theory the critical exponents of any $d$-dimensional $O(n)$
random-field model (RFM) (for which the Kramers degeneracy is broken by
the randomness) appear to be identical to those of an ordinary $O(n)$
model of dimension $d - 2$. For the Ising ($n = 1$) case, this
dimensional reduction was shown rigorously to be
incorrect.\cite{Imb84,BK87}  Another interesting development was the
calculation of Mezard and Young,\cite{MY92} who showed that random
fields caused breaking of replica symmetry below $T_c$ for any
finite value of $n$.  Thus there is no good reason to expect that
dimensional reduction should be correct near $T_c$ for any finite
value of $n$.

Although there is certainly a family resemblance between the RFM and
the RAM, the difference between breaking the Kramers degeneracy at
the level of Hamiltonian and breaking it spontaneously has profound
consequences.  One such consequence is Theorem 4.4 of Aizenman and
Wehr,\cite{AW90} which applies to the RFM, but not to the RAM.  A
naive but not entirely misleading analogy may be drawn between the
relationship of the RFM to the RAM and the relationship between
applying a uniform magnetic field to a ferromagnet or to an
antiferromagnet.  The field which couples linearly to the order
parameter has a qualitatively stronger effect than the field which
couples quadratically to the order parameter.

Translation invariance of $H_{RAM}$ is broken for any non-zero value
of $D$, since the vectors $\hat{\bf n}_{i}$ are random.  Within a
high-temperature perturbation theory, performing a configuration average
over the ensemble of random lattices appears to restore translation
invariance above $T_c$.  However, the radius of convergence of this
perturbation theory cannot be greater for $D \ne 0$ than it is for
$D = 0$.  For models described by Eqn.~(1), the $T_c$ predicted by
extrapolating the low orders of perturbation theory is always maximal at
$D = 0$.  This implies that for $D \ne 0$ the high-temperature
perturbation theory does not converge near $T_c$.  The inadequacy of
perturbation theory to describe $XY$ models in $d = 3$, because of the
effects of vortex lines, has been discussed by Halperin.\cite{Hal81}
While it is not clear than Halperin's argument is valid for the RFM,
where the Kramers degeneracy is broken by the Hamiltonian, it should be
valid for the RAM.  Thus it seems quite implausible that for $d < 4$
the twist energy for Eqn.~(2) really scales as $L^{d - 2}$ when $D \ne 0$,
even though this is correct to all orders in the configuration-averaged
perturbation theory.

The argument of Pelcovits\cite{Pel79} for the $n = 2$ RAM, which is a
prototype for much subsequent work,\cite{Fish85} assumes that if one goes
to small enough $D / J$ and low enough $T$ then the effects of vortex
lines can be ignored.  In essence, what is done is to replace the spin
variables by a noncompact "elastic manifold".  These authors then claim
that this does not effect the behavior one is studying.  However, this
cannot be true when one considers behavior on scales larger than the
Imry-Ma length.\cite{Cop91}

The basic point is that Imry-Ma-type arguments for continuous O({\it n})
spins ({\it i.e.} $n \ge 2$) are not self-consistent.  One begins by
assuming that the random field is weak so that the twist energy scales
as $L^{d - 2}$, as in the absence of the randomness.  Then one shows
that, if $d \le 4$ and $T < T_c$, the effective coupling to the random
field increases as $L$ increases.  If the effective random coupling is
strong, however, then assuming that the twist energy is uniformly
distributed throughout the volume is not reasonable.  The conclusion
which should be drawn from this is that a deeper analysis is needed when
$d \le 4$.

In order to understand whether the problems with perturbation theory are
actually due to vortex lines, and thus restricted to the $n = 2$ case, or
if similar problems can also be expected for $n > 2$, it may be helpful
to reconsider the analysis of Pelcovits, Pytte and Rudnick.\cite{PPR78}
These authors show that within their perturbation theory the pure O({\it n})
mean-field theory critical fixed point remains stable against random
anisotropy for $d > 4$.  This contrasts to the random-field case, where
mean-field theory is only stable for $d > 6$.  Then they argue that for
$d \le 4$ and $n \ge 2$ there is no stable critical fixed point for the
RAM, because under rescaling transformations the random anisotropy coupling
constant runs off to $\infty$.  However, they did not (and within their
formulation could not) examine the possibility that there could exist another
ferromagnetic critical fixed point at a large value of $D / J$.  The reason
why such an object may exist is that there exist alternative formulations of
mean-field theory\cite{DV80,HCB87} for the RAM in the limit $D / J \to \infty$.

It is useful to consider the generalization of Eqn.~(2) to $p$-fold random
fields.\cite{Aha81}  In the $d = 2$ case\cite{HKY81,CO82} it has been shown
that there continues to be a Kosterlitz-Thouless phase as long as $p^2 > 8$,
{\it i.e.} for $p^2 > 8$ a weak $p$-fold random field does not destroy the
Kosterlitz-Thouless phase.  It was claimed by Aharony\cite{Aha81} that
ferromagnetism should be unstable for any value of $p$ when $d = 3$.  However,
a computer simulation study\cite{Fis92} for $p = 3$ is not consistent with
this claim, which is based on the weak randomness perturbation theory around
the $D = 0$ model.  This $d = 3$ computer simulation finds that there is a mass
gap at $T = 0$, an effect which cannot be reproduced within the perturbation
theory.  The interpretation of this is that for $p \ge 3$ the thickness of a
domain wall remains finite in the limit $L \to \infty$, {\it i.e.} the domain
wall becomes localized by random pinning.

Removing vortex lines from the pure $XY$ model by letting the vortex fugacity
become large forces the system into a ferromagnetic state at any
temperature.\cite{KSW86,Fis95b}  This result is true even in the presence of
a strong $p = 2$ random anisotropy,\cite{Fis00} but the $p = 2$ case is more
complicated than $p \ge 3$.  For $p = 2$, as we shall see, the domain walls
probably have a fractal structure, rather than becoming completely localized.

\section{Numerical results}

In this work, we will present results obtained from heat bath Monte Carlo
calculations.  The data were obtained from $L \times L \times L$ simple cubic
lattices with $16 \le L \le 64$ using periodic boundary conditions.  The
calculations were done for a 12-state clock model, {\it i.e.} a ${\bf Z}_{12}$
approximation\cite{Fis97} to the $XY$ model of Eqn.~(2).  The computer program
was an adaptation of the code used recently for the $XY$ model in a random
field,\cite{Fis07} modified to replace the random field term with the random
2-fold anisotropy term of Eqn.~(2).  For any integer value of the quantity
$D / 2 J$ one can use a lookup table for the Boltzmann factors, because all
the energies in the problem are then expressible as sums of integers and
integer multiples of $\sqrt{3}$.  The values of $D / 2 J$ for which data were
obtained are 1, 2, 3, 6, and $\infty$.

The discretization of the phase space of the model has significant effects at
very low $T$, but the effects at the temperatures we study here are expected
to be negligible compared to our statistical errors.  The probability
distributions for the local magnetization of equilibrium states which are
calculated for the ${\bf Z}_{12}$ model are found to have very small
contributions from the third and higher harmonics of $\cos ( \phi )$ and
$\sin ( \phi )$.  This is strong evidence that the 12-state clock model is
an accurate approximation to the $XY$ model within our range of
parameters.  The ${\bf Z}_{12}$ model shows equivalent behavior for $D$ and
$- D$, unlike the ${\bf Z}_{6}$ model used earlier.\cite{Fis95}

The program uses two independent linear congruential pseudorandom number
generators, one for choosing the values of the $\theta_i$, and a different one
for the Monte Carlo spin flips, which are performed by a single-spin-flip
heat-bath algorithm.  The code was checked by setting $D = 0$, and seeing that
the known behavior of the pure ferromagnetic system was reproduced correctly.

Each sample was started off in a random spin state at a temperature
significantly above the $T_c$ for the pure model, and cooled slowly.  Thermal
averages for $S (\vec{\bf k})$ were obtained at a set of temperatures spanning
the critical region.

\begin{figure}
\includegraphics[width=3.4in]{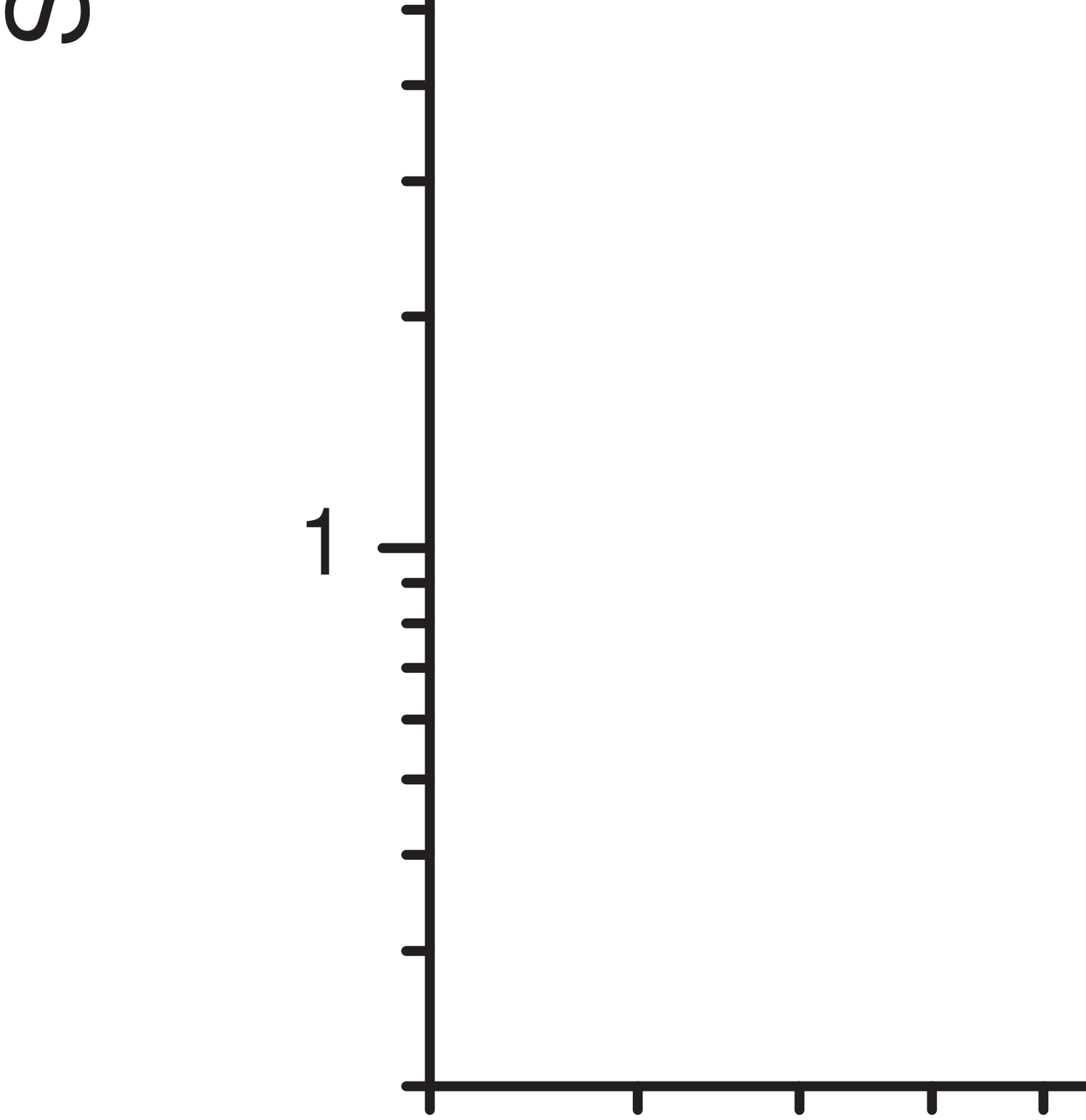}
\caption{\label{Fig.1} Angle-averaged structure factor for $64 \times 64 \times 64$
lattices with $D / 2 J = 1$ at $T = 2.203125$. The axes are scaled logarithmically.}
\end{figure}

\begin{figure}
\includegraphics[width=3.4in]{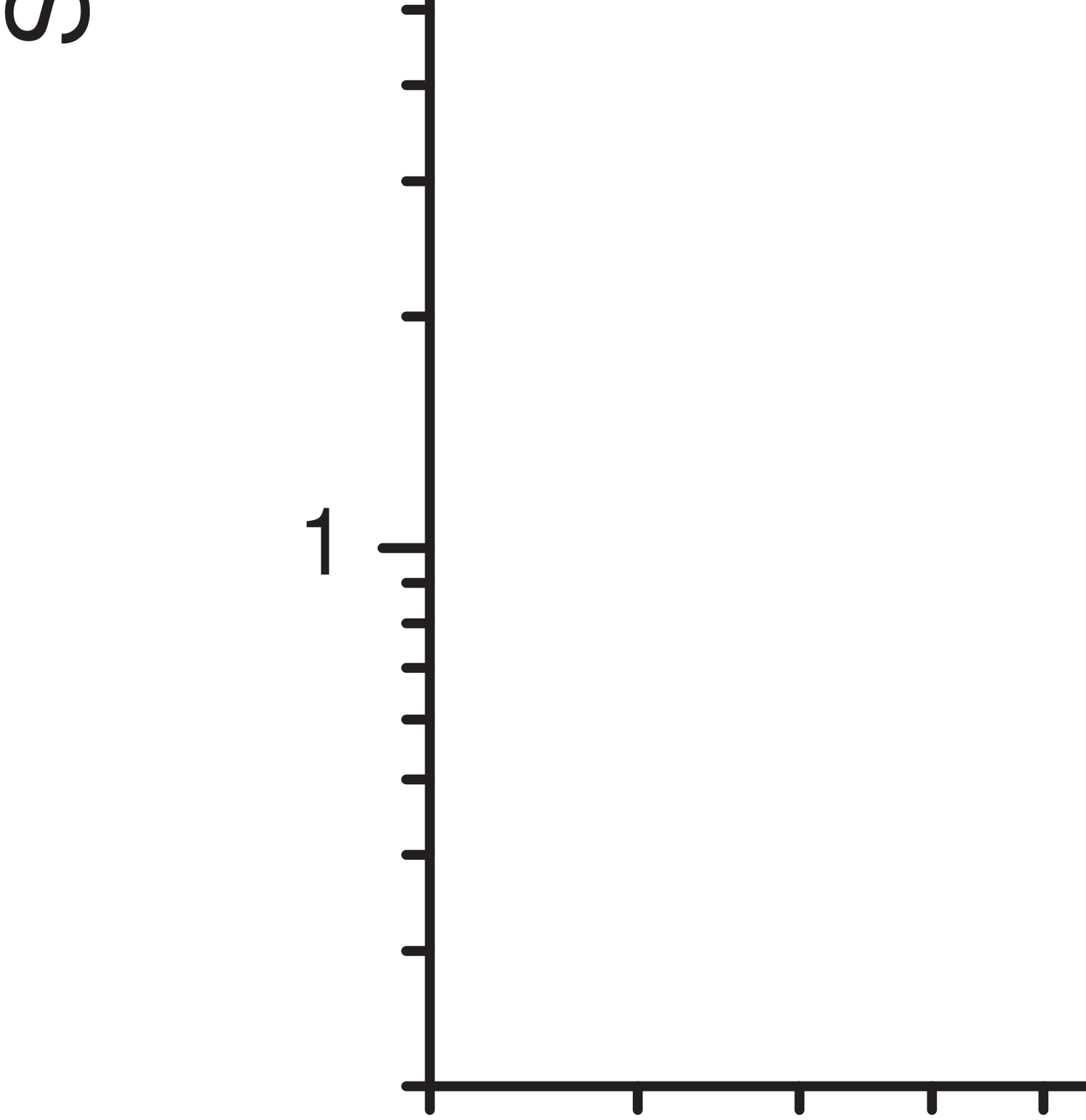}
\caption{\label{Fig.2} Angle-averaged structure factor for $64 \times 64 \times 64$
lattices with $D / 2 J = 2$ at $T = 2.1875$. The axes are scaled logarithmically.}
\end{figure}

\begin{figure}
\includegraphics[width=3.4in]{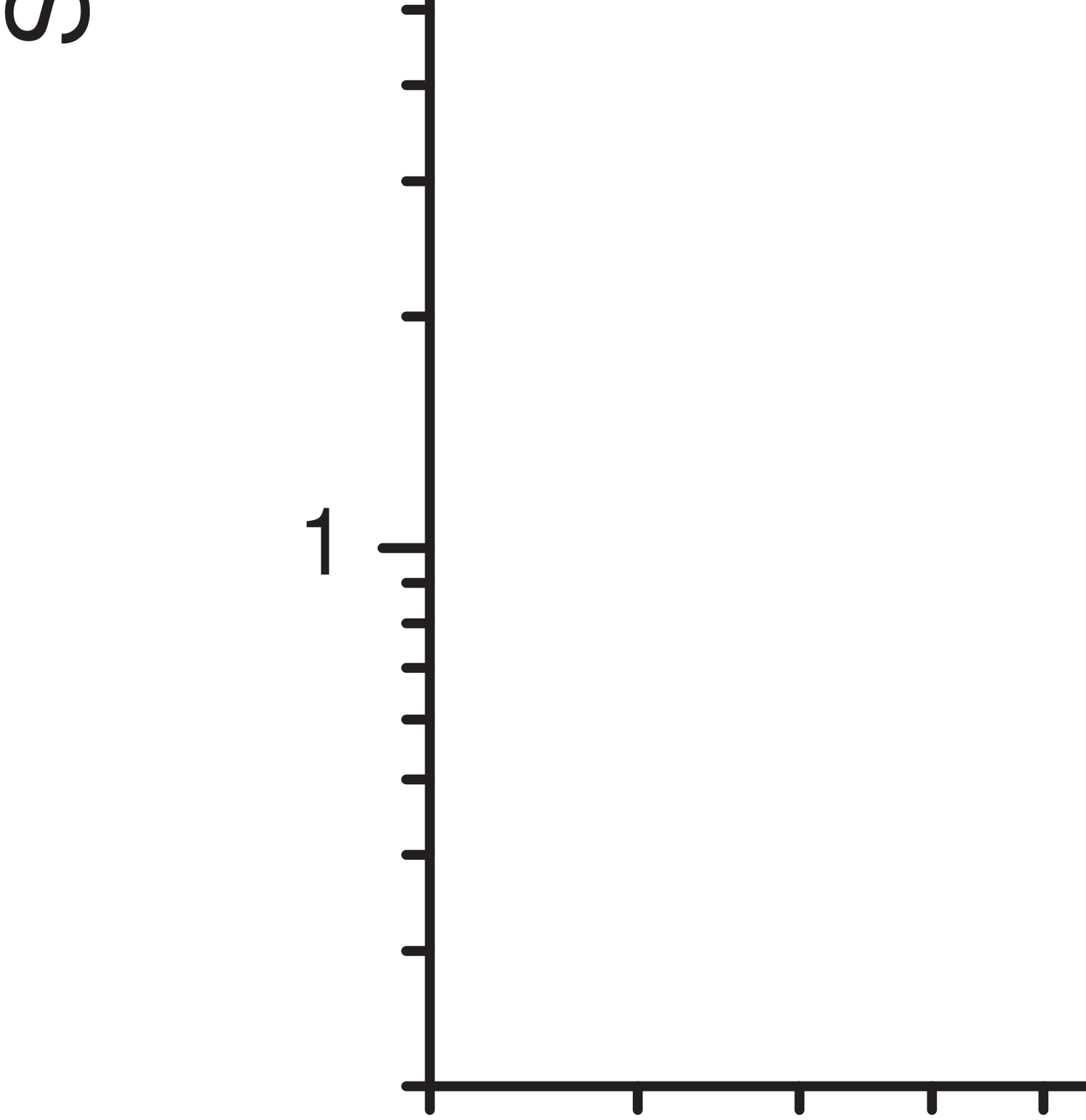}
\caption{\label{Fig.3} Angle-averaged structure factor for $64 \times 64 \times 64$
lattices with $D / 2 J = 3$ at $T = 2.171875$. The axes are scaled logarithmically.}
\end{figure}

\begin{figure}
\includegraphics[width=3.4in]{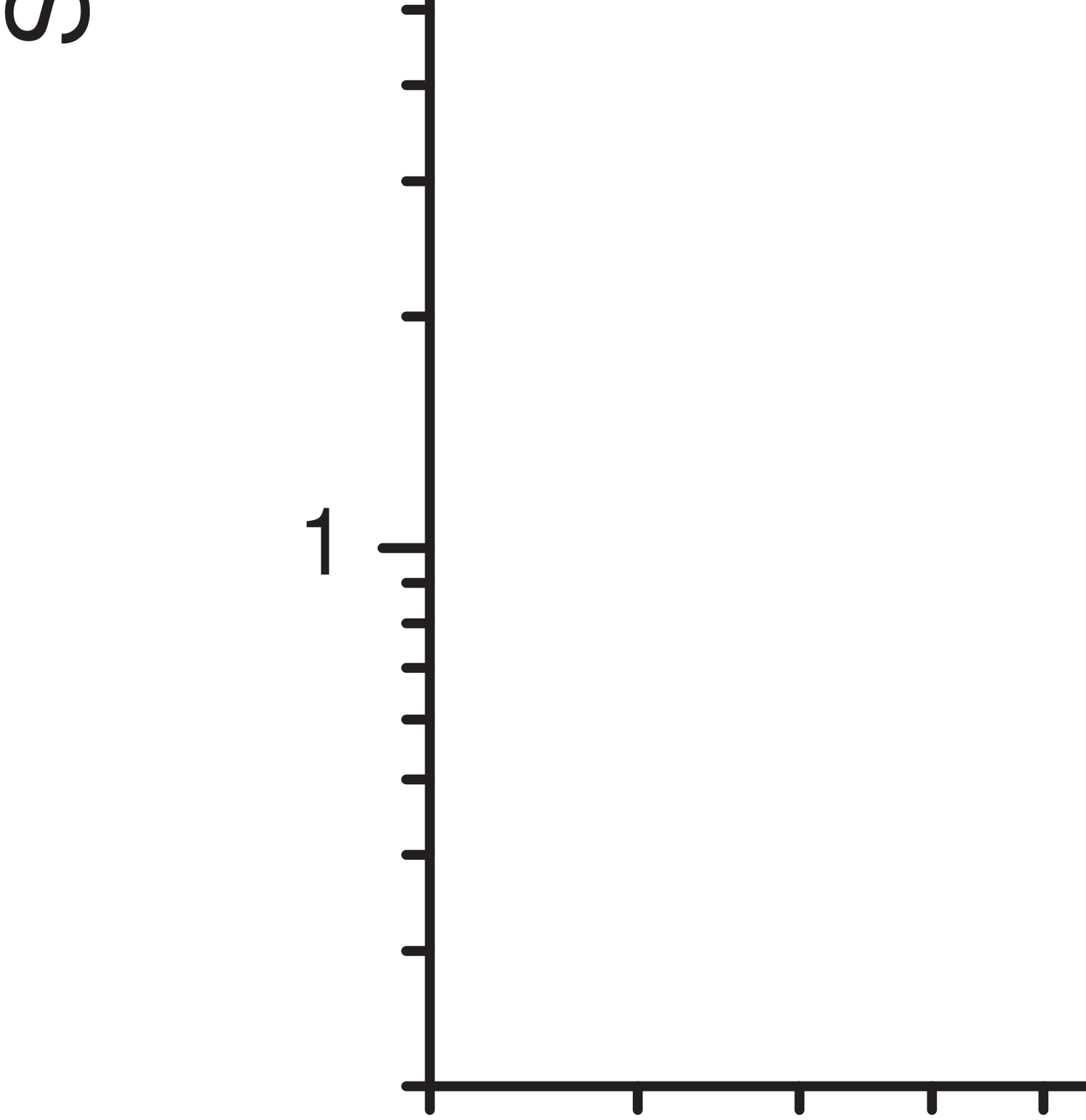}
\caption{\label{Fig.4} Angle-averaged structure factor for $64 \times 64 \times 64$
lattices with $D / 2 J = 6$ at $T = 2.078125$. The axes are scaled logarithmically.}
\end{figure}

\begin{figure}
\includegraphics[width=3.4in]{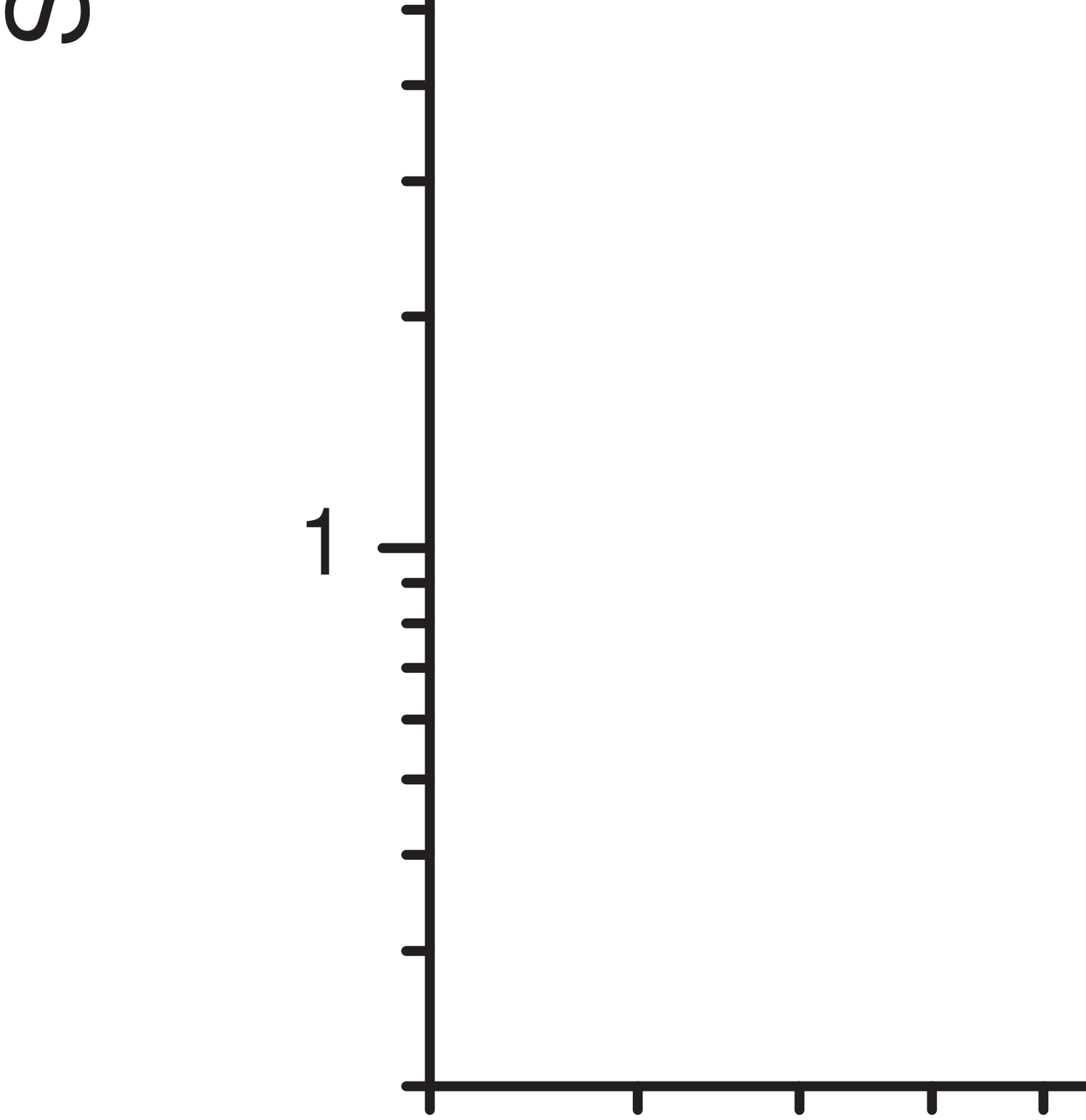}
\caption{\label{Fig.5} Angle-averaged structure factor for $64 \times 64 \times 64$
lattices with $D = \infty$ at $T = 1.921875$. The axes are scaled logarithmically.}
\end{figure}

The magnetic structure factor, $S (\vec{\bf k}) = \langle | \vec{\bf M}(\vec{\bf
k}) |^2 \rangle $, for $n = 2$ spins is
\begin{equation}
  S (\vec{\bf k}) ~=~  L^{-3} \sum_{ i,j } \cos ( \vec{\bf k} \cdot
  \vec{\bf r}_{ij}) \langle \cos ( \phi_{i} - \phi_{j}) \rangle  \,   ,
\end{equation}
where $\vec{\bf r}_{ij}$ is the vector on the lattice which starts at
site $i$ and ends at site $j$.  Here the angle brackets denote a thermal
average.  For a RAM with $n > 1$, unlike the RBIM, the longitudinal
part of the magnetic susceptibility, $\chi_l$, which is given by
\begin{equation}
  T \chi_l (\vec{\bf k}) ~=~ 1 - M^2 ~+~ L^{-3} \sum_{ i \ne j } \cos (
  \vec{\bf k}  \cdot \vec{\bf r}_{ij}) (\langle \cos ( \phi_{i} - \phi_{j}
  ) \rangle ~-~ Q_{ij} )  \,   ,
\end{equation}
where $M^2 ~=~ \langle |\vec{\bf M}| \rangle^2$, and $Q_{ij} ~=~ \langle
\vec{\bf S}_{i} \rangle \cdot \langle \vec{\bf S}_{j} \rangle$.  For O(2)
spins
\begin{equation}
  M^2 ~=~ L^{-6} [ \langle |\sum_{i} \cos ( \phi_{i} )| \rangle^2
      ~+~ \langle |\sum_{i} \sin ( \phi_{i} )| \rangle^2 ] \,  ,
\end{equation}
and
\begin{equation}
  Q_{ij} ~=~ \langle \cos ( \phi_{i} ) \rangle \langle \cos ( \phi_{j} )
  \rangle ~+~ \langle \sin ( \phi_{i} ) \rangle \langle \sin ( \phi_{j} )
  \rangle  \,  .
\end{equation}

Thus $M^2$ is not the same as $S$, even above $T_c$.  The scalar
quantity $M^2$ is a well-behaved function of the lattice size $L$ for
finite lattices, which approaches its large $L$ limit smoothly as
$L$ increases, except possibly at a phase transition. The vector
$\vec{\bf M}$, on the other hand, may not be a well-behaved function
of $L$ for an $XY$ model in a two-fold random field. Knowing the local
direction in which $\vec{\bf M}$ is pointing, averaged over some small
part of the lattice, may not give us a strong constraint on what
$\vec{\bf M}$ for the entire lattice will be.

The critical exponent $\eta$, is defined at $T = T_c$ by the small
$\vec{k}$ behavior
\begin{equation}
  S (\vec{\bf k}) ~\approx~ c |\vec{\bf k}|^{- (2 - \eta)}   \,  ,
\end{equation}
where $c$ is some constant.  For each value of $D / J$, results for four
different $L = 64$ configurations of the random anisotropy $\theta_i$ were
averaged.  The same four samples of random $\theta_i$ were used for all
values of $T$, and all values of $D / J$, in order to facilitate the
comparison of results for different values of $T$ and $D$.

All of the data shown in these figures were obtained from Monte Carlo runs
which used hot start initial conditions, starting at at temperature well
above $T_c$.  The value of $T$ was then lowered in steps.  The initial part
of the run at each $T$ was discarded to allow the system to equilibrate.  For
these $L = 64$ runs with $D / 2 J$ = 1, at each $T$ a sequence of spin states
obtained at intervals of 20,480 Monte Carlo steps per spin (MCS) was Fourier
transformed and averaged. For the larger values of $D$, where the relaxation
times are longer, this interval was chosen to be 102,400 MCS.  The number of
these selected spin states was chosen to be 16 for each of the finite values
of $D / J$, and 32 for $D / J = \infty$.  The Fourier transformed spin state
data were then binned according to the values of $k$ = $|\vec{\bf k}|$, to
give the angle-averaged $S ( k )$.  Finally, a configuration average over
the four random samples was performed.  Both equally weighted and
logarithmically weighted averages were tried.  No significant differences
were found between these two types of weighting, and only the equally
weighted averages will be displayed here.

The results for $D / 2 J$ = 1, 2, 3, 6, and $\infty$ are shown in Figs.~1,
2, 3, 4, and 5, respectively.  The values of $T$ which are used in these
figures are convenient binary fraction approximations to the values of $T_c$
at these values of $D / 2 J$.  The best estimates of $T_c$ were determined
later, by the analysis of the data over a range of $L$ and $T$.  We see from
these figures that $S ( k )$ is only a weakly varying function of $D / J$,
at least for $L = 64$.

The values of $2 - \eta$, as displayed on the figures, were found by
least-squares fits to the data points for $0 < k < \pi/8$, where the data
are well-approximated by Eqn.~8.  Note that $\eta$ appears to be
a slowly varying monotonic function of $D / J$, and that the extrapolation of
$\eta$ down toward $D$ = 0 appears to be significantly different from the
value of $\eta$ found for the nonrandom $n = 2$
ferromagnet.\cite{LGZJ80,LT89}  It is also interesting to note that the
value of $\eta$ found for $D = \infty$ appears to be identical to the value
of $\eta$ for the nonrandom system, but the significance of this is unclear.

The fact that $\eta$ appears to vary with $D / J$ is an indication that the
claim of Reed\cite{Reed91} is too simplistic.  He did not calculate a
numerical value for $\eta$, but he argued that the finite-size scaling (FSS)
behavior at $D / 2 J$ = 1 was indistinguishable from that of the nonrandom
system.

One should not conclude from these data that $\eta$ is varying continuously with
$D$, so that there is a line of critical points.  Another explanation of the data
is that for any $D$ we have a function $D_{eff} ( D / J, L )$, which increases
very slowly as $L$ increases, up to a value $D_{eff} = D^*$  Then we will only
find $\eta_{eff} = \eta^*$ when $L$ becomes large enough so that $D_{eff}
\approx D^*$.  In the Cayley tree mean-field approximation,\cite{HCB87} whether
$D^*$ is finite or infinite depends on the value of $z / n$.

\begin{figure}
\includegraphics[width=3.1in]{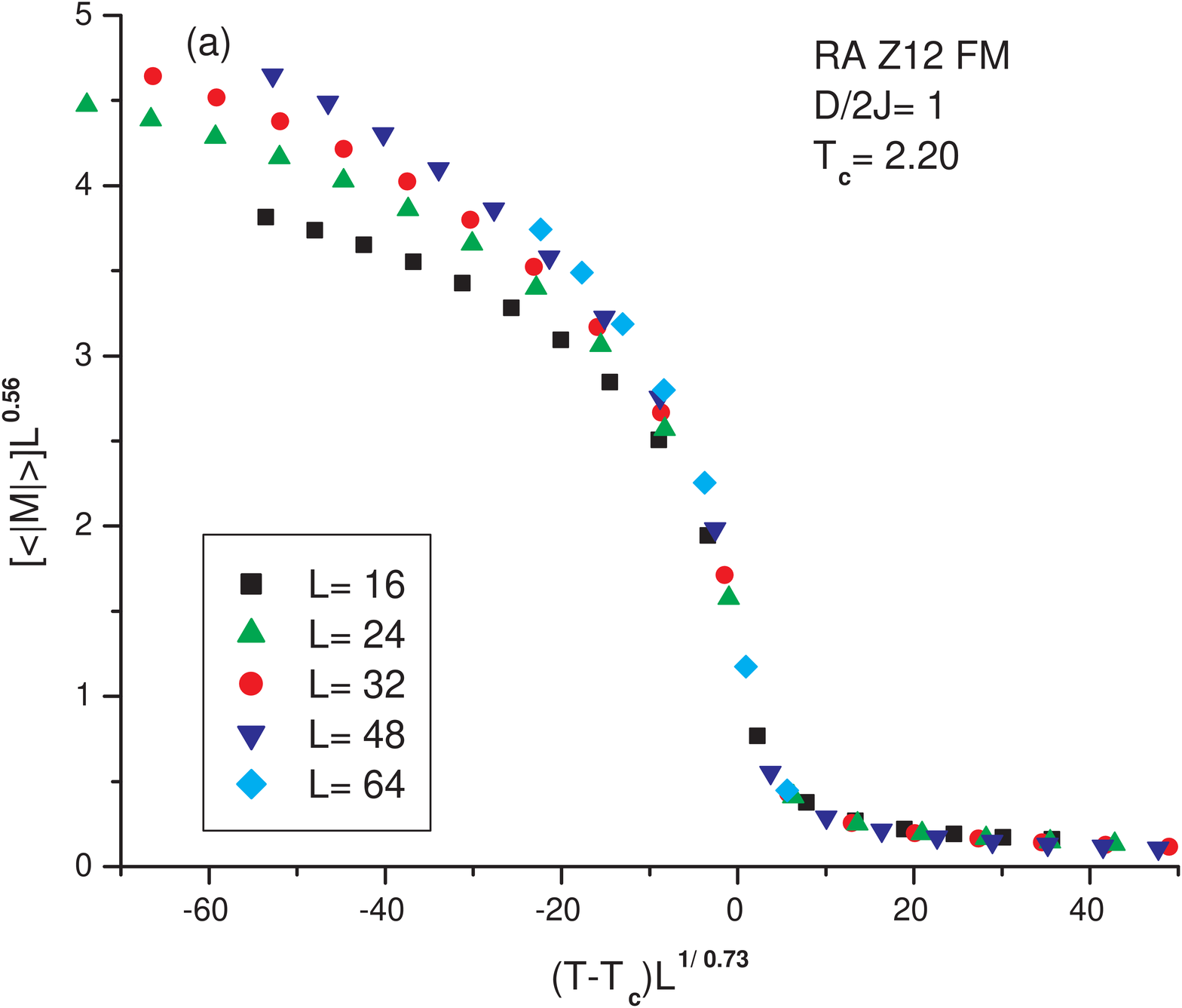}\quad
\includegraphics[width=3.1in]{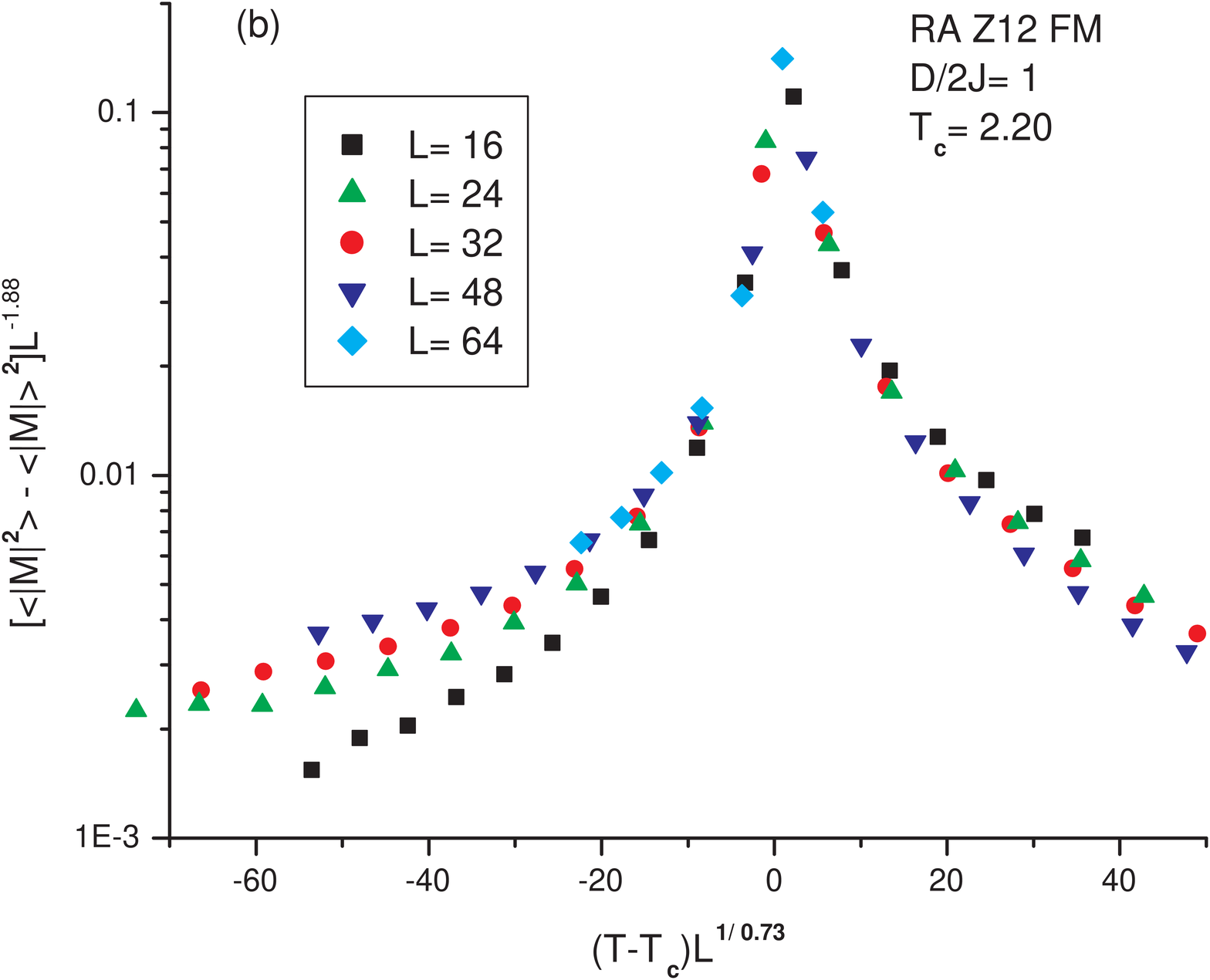}
\caption{\label{Fig.6}(color online) Finite-size scaling near $T_c$ for
$L \times L \times L$ lattices with $D / 2 J = 1$.
(a) Configuration-averaged magnetization vs. temperature.
(b) $\chi_l$ vs. temperature.  The $y$-axis is scaled logarithmically.}
\end{figure}

\begin{figure}
\includegraphics[width=3.1in]{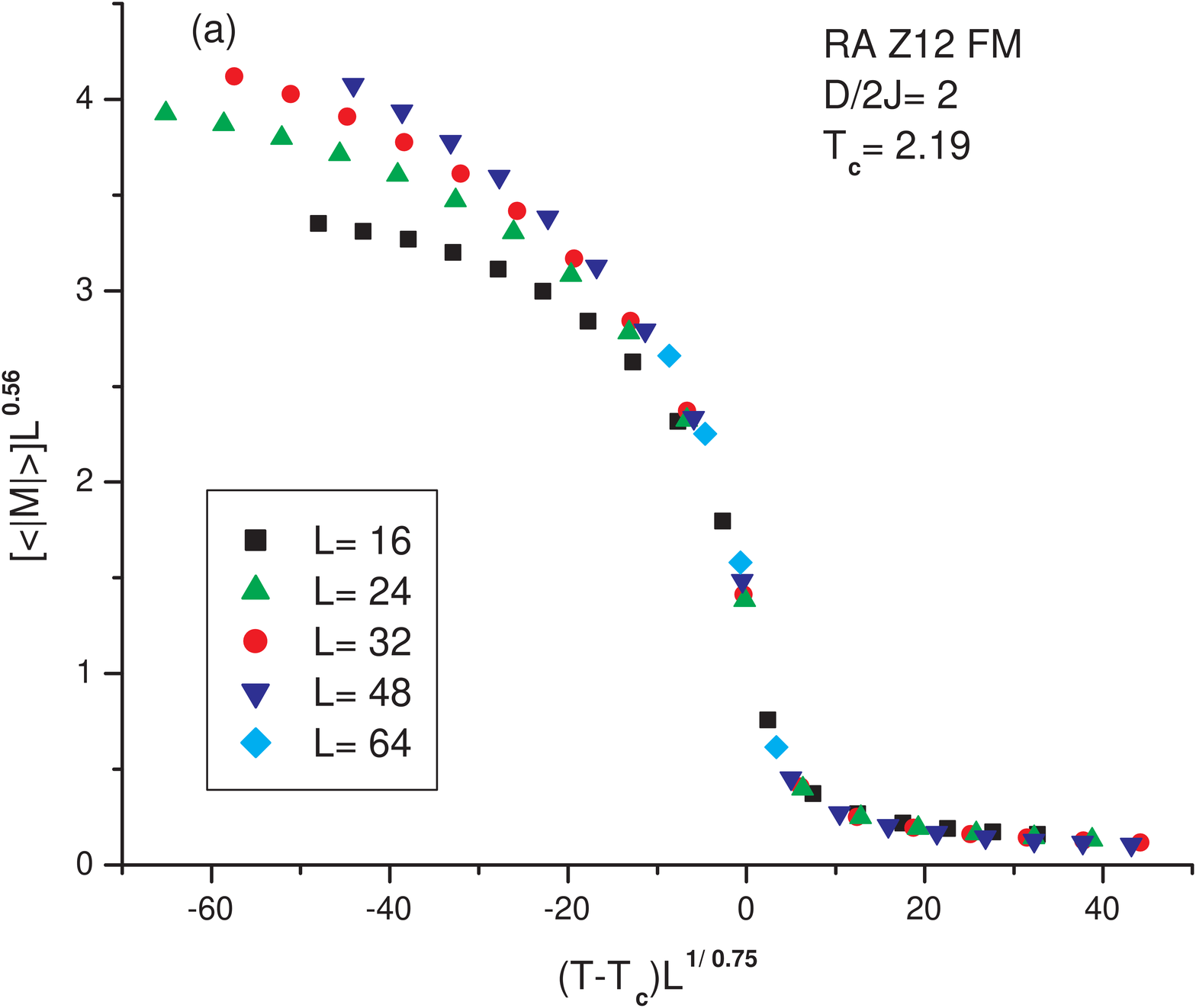}\quad
\includegraphics[width=3.1in]{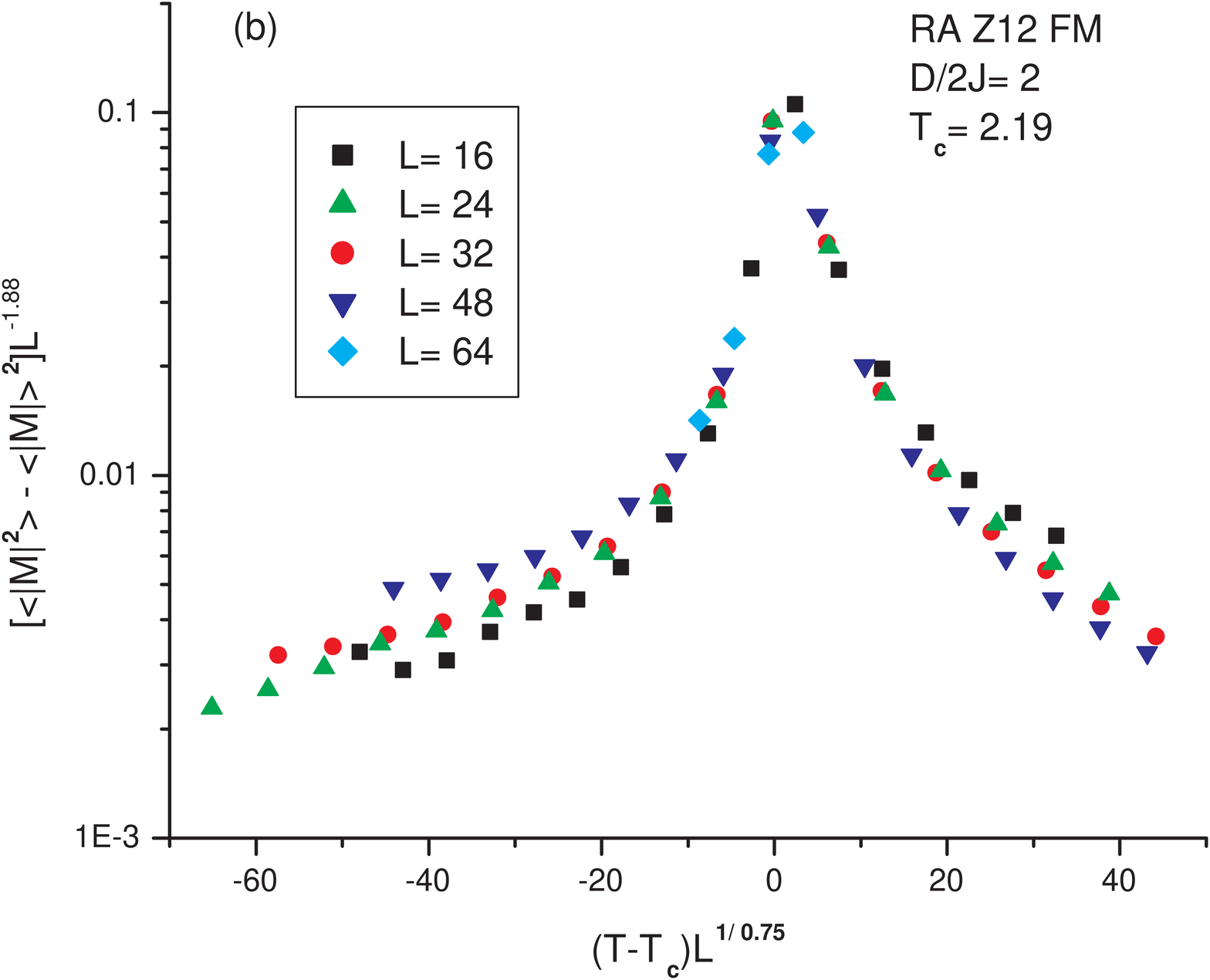}
\caption{\label{Fig.7}(color online) Finite-size scaling near $T_c$ for
$L \times L \times L$ lattices with $D / 2 J = 2$.
(a) Configuration-averaged magnetization vs. temperature.
(b) $\chi_l$ vs. temperature.  The $y$-axis is scaled logarithmically.}
\end{figure}

\begin{figure}
\includegraphics[width=3.1in]{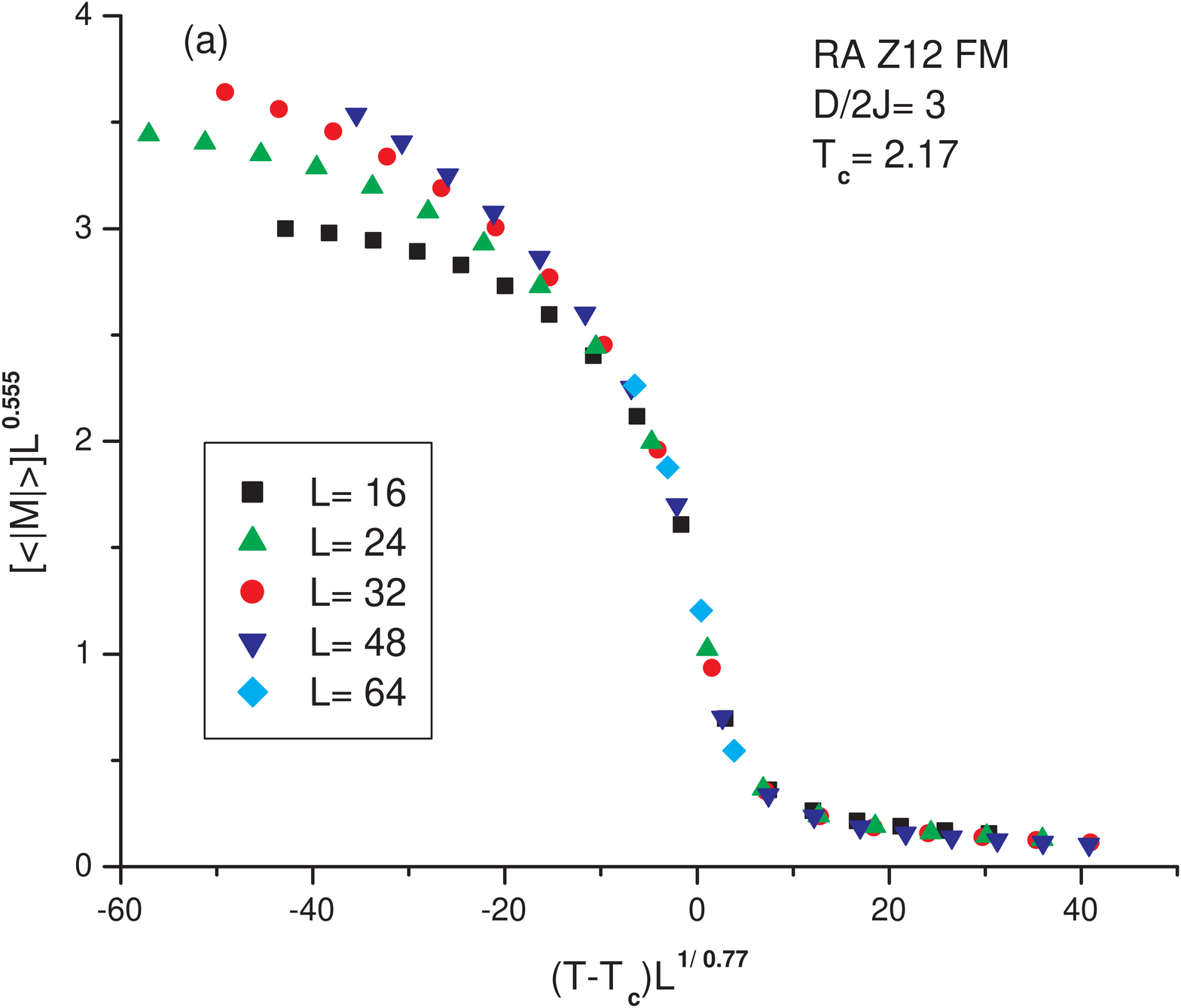}\quad
\includegraphics[width=3.1in]{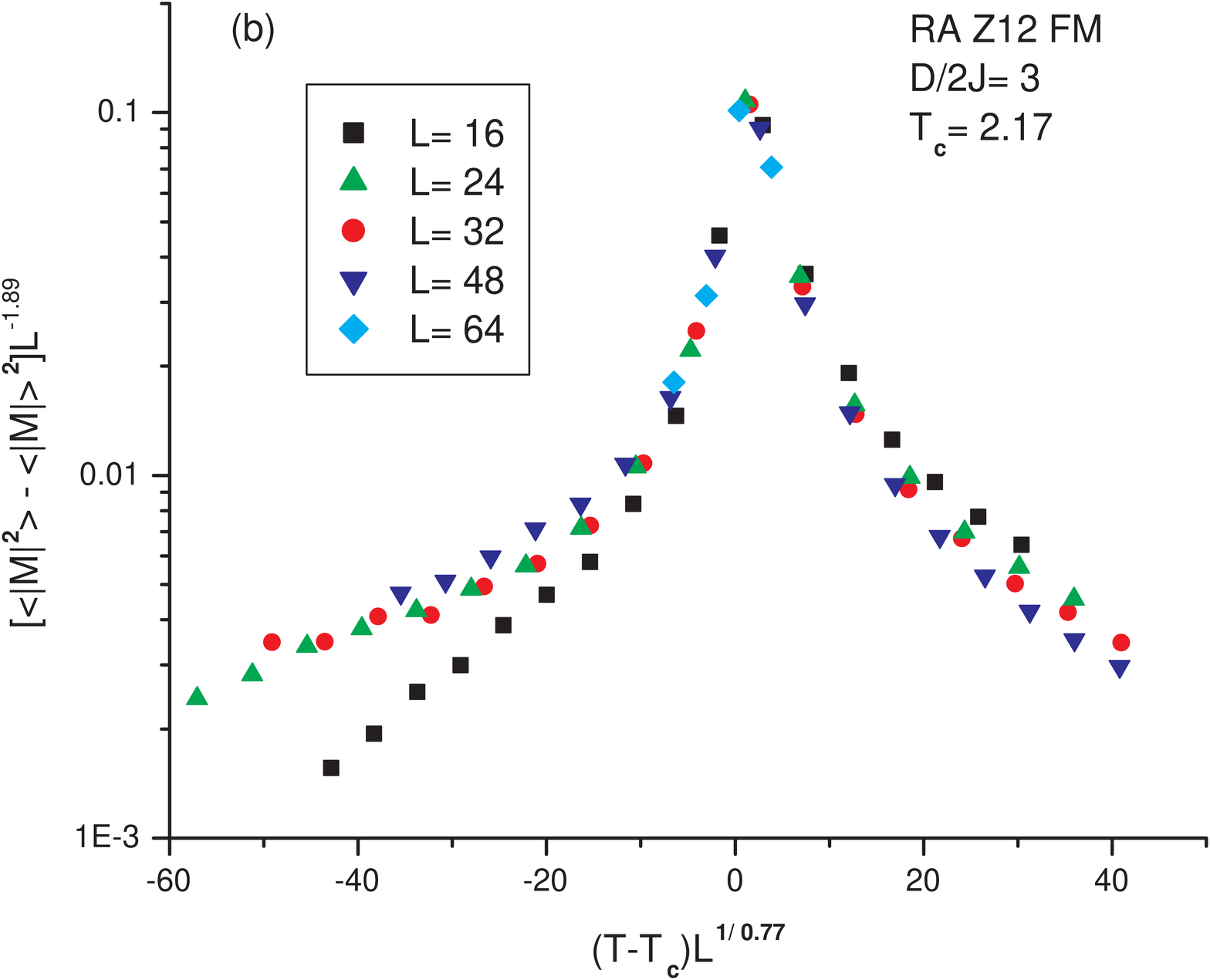}
\caption{\label{Fig.8}(color online) Finite-size scaling near $T_c$ for
$L \times L \times L$ lattices with $D / 2 J = 3$.
(a) Configuration-averaged magnetization vs. temperature.
(b) $\chi_l$ vs. temperature.  The $y$-axis is scaled logarithmically.}
\end{figure}

\begin{figure}
\includegraphics[width=3.1in]{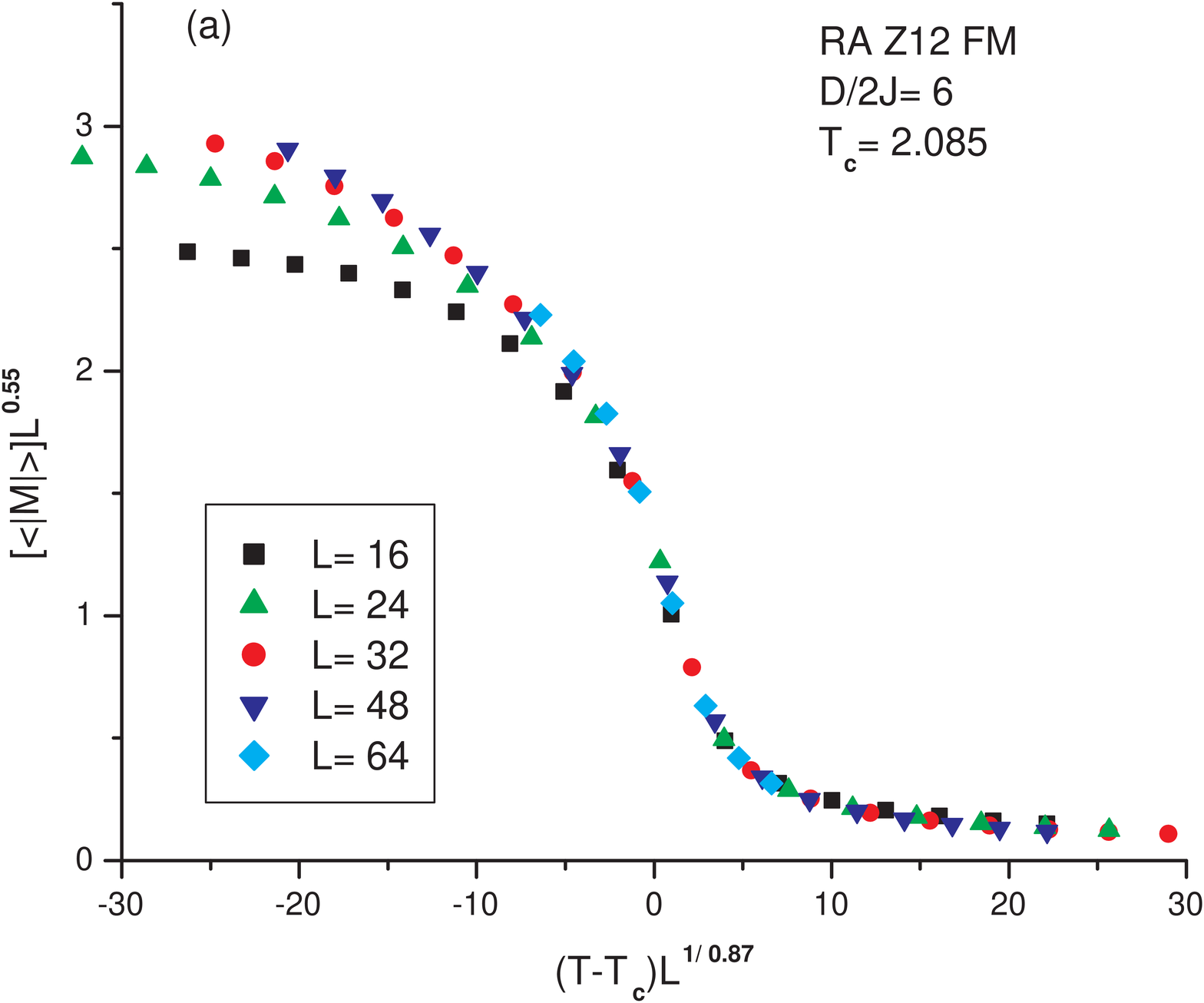}\quad
\includegraphics[width=3.1in]{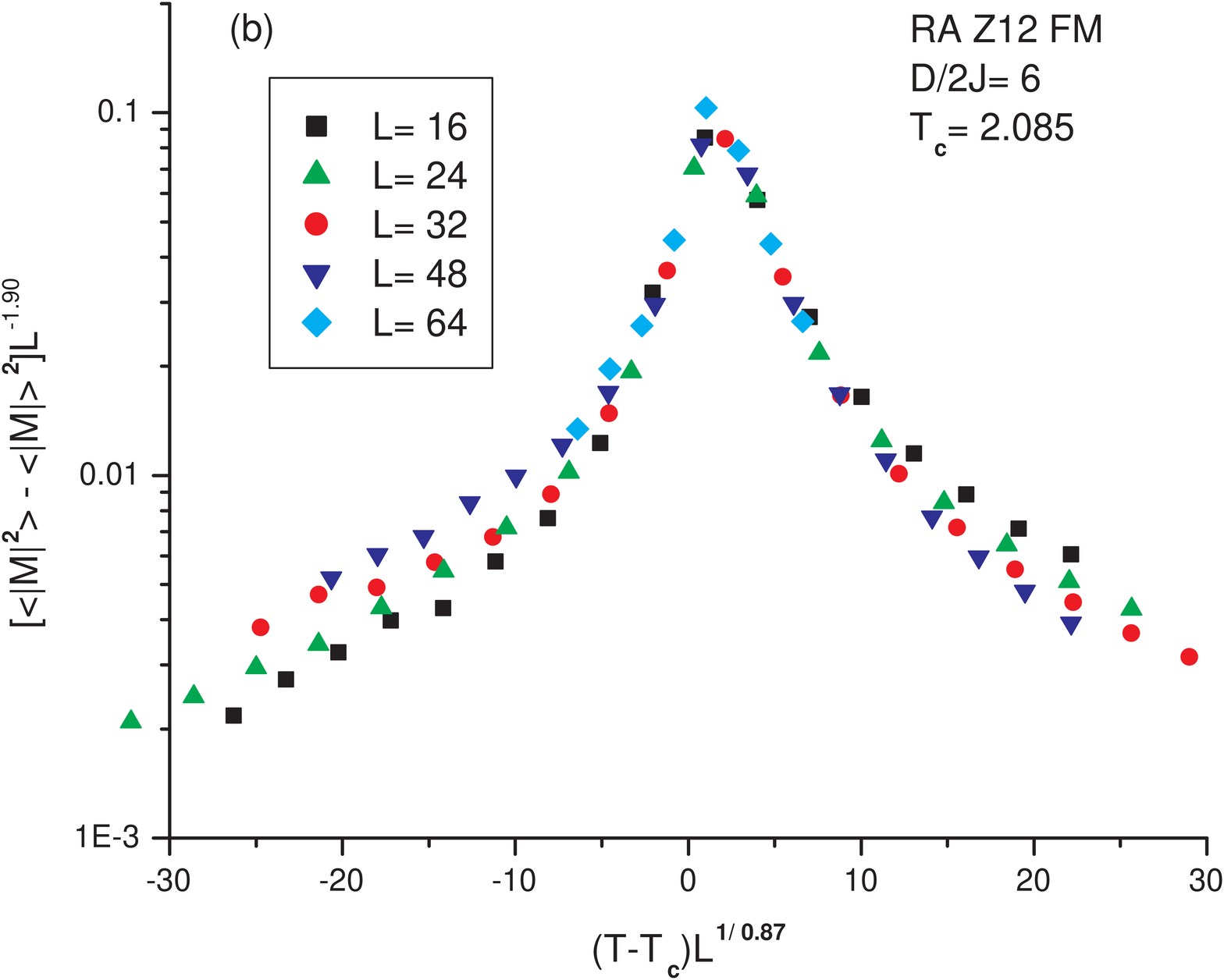}
\caption{\label{Fig.9}(color online) Finite-size scaling near $T_c$ for
$L \times L \times L$ lattices with $D / 2 J = 6$.
(a) Configuration-averaged magnetization vs. temperature.
(b) $\chi_l$ vs. temperature.  The $y$-axis is scaled logarithmically.}
\end{figure}

\begin{figure}
\includegraphics[width=3.1in]{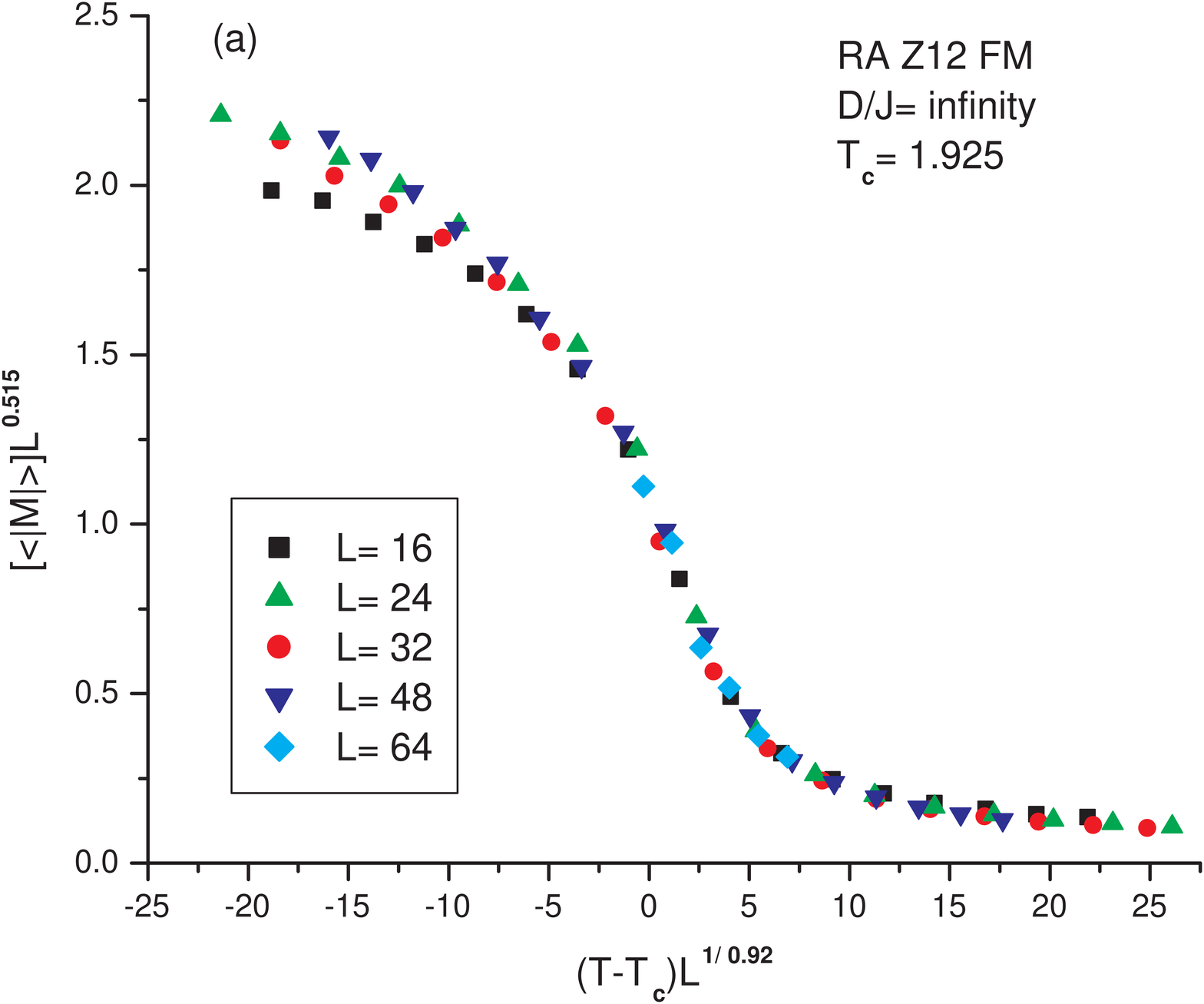}\quad
\includegraphics[width=3.1in]{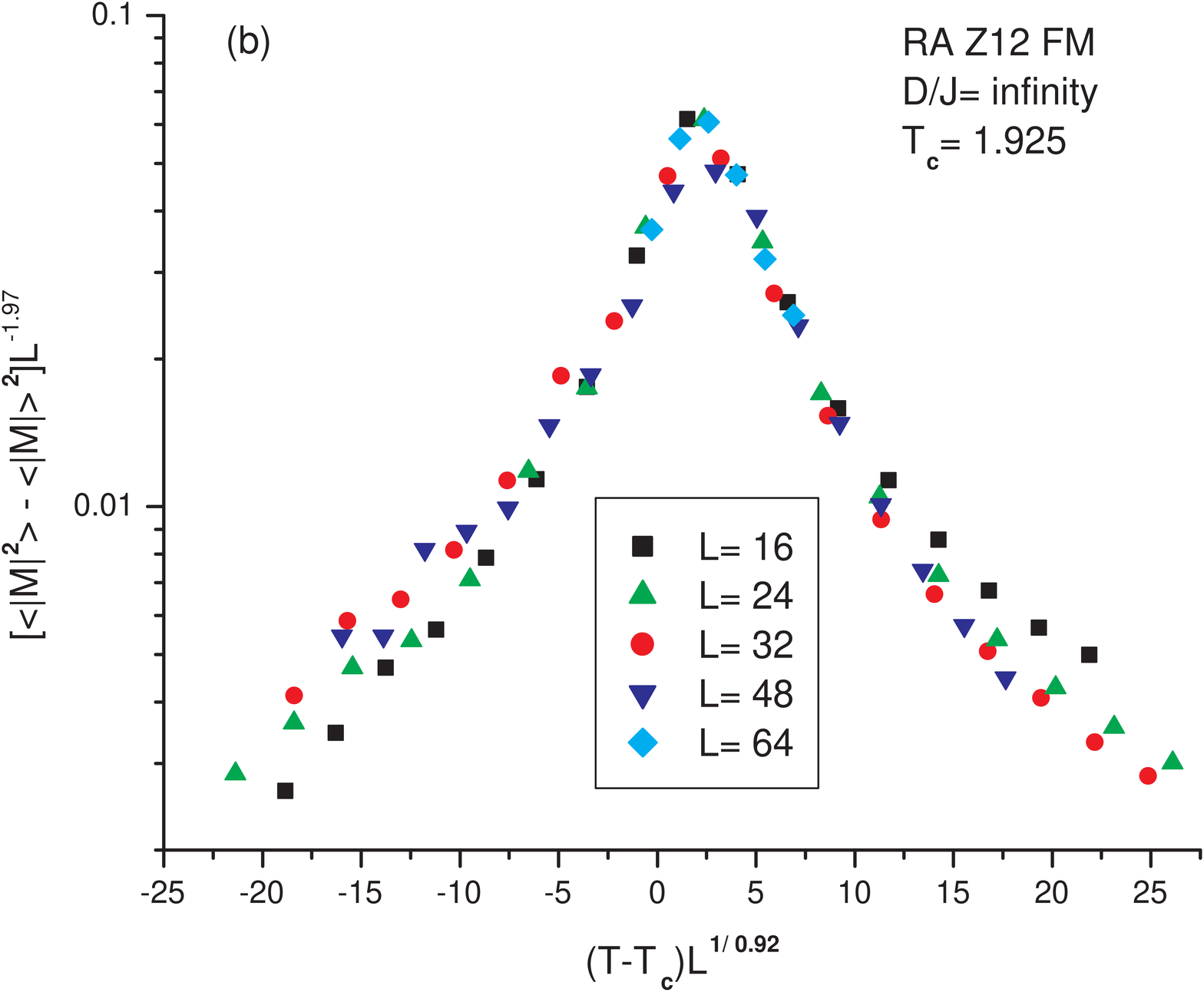}
\caption{\label{Fig.10}(color online) Finite-size scaling near $T_c$ for
$L \times L \times L$ lattices with $D / J = \infty$.
(a) Configuration-averaged magnetization vs. temperature.
(b) $\chi_l$ vs. temperature.  The $y$-axis is scaled logarithmically.}
\end{figure}

If we make the assumption that the usual critical exponent scaling laws
for translation invariant models remain valid for the RAM, we can easily
obtain values of the exponent combinations $\beta / \nu$ and $\gamma / \nu$
from our computed values of $2 - \eta$.  These combinations are exactly
what we need for FSS of the magnetization $| \vec{\bf M} ( L, T ) |$ and the
magnetic susceptibility\cite{FB72} $\chi_l ( L, T )$.  Thus, by making
standard FSS plots,\cite{NB99} we can test the validity of these scaling
laws for the RAM.

In Fig.~6(a) we show a FSS plot of the configuration average of $| \vec{\bf M}
( L, T ) |$ on $L \times L \times L$ lattices, for $L$ between 16 and 64.  The
number of sample configurations used for each $L < 64$ was 8 for $D / 2 J$ =
1, 2, 3 and 6, and 16 for $D = \infty$.  For $L$ = 64, the number of samples
was 4 for all $D$.  Fig.~6(b) shows a similar plot for $\chi_l$.  Figs.~7, 8,
9 and 10 show the corresponding plots for $D / 2 J$ = 2, 3, 6 and $\infty$,
respectively.  Since the values of $\eta$ used here were taken from the fits to
the small k behavior of $S ( k )$, the only two adjustable fitting parameters
used in these figures were the values of $\nu$ and $T_c$, which were required
to be identical for parts (a) and (b) of each figure.

In these FSS plots, the temperature coordinate scales as $(T - T_c)L^{1 /
\nu}$.  The reader should note that the range of $T$ which we cover in these
plots is about an order of magnitude larger than the range which one would
typically use for a problem where one is already confident about the nature
of the phase transition, and one is trying to obtain high precision estimates
of $T_c$ and the critical exponents by concentrating on the range of $T$ where
$\xi \approx L$.  As a consequence of this, the spacings between the values
of $T$ for which we have taken data are rather large.  Thus we are unable to
use histogram reweighting\cite{FLS95} to obtain essentially continuous values
for the thermodynamic functions.

From the results given in these figures, we see that the estimates of $\nu$
increase monotonically and the estimates of $T_c$ decrease monotonically as
$D / 2 J$ increases.  We also see that the peak in $\chi_l$ is slightly
above $T_c$ for finite $L$, which is typical for ferromagnetic critical
behavior.  The data collapse is good near this peak, which is the range of $T$
for which $\xi > L$.  We do not give estimates of statistical errors for $\nu$,
because  we believe that the variation in $\nu$ in the range $D / 2 J$ = 1 to 6
is due to variation in the value of $D_{eff}$.  We will discuss this further in
the next section.  The errors in the values of $T_c$ are estimated to be less
than $\pm 0.01$.

\begin{figure}
\includegraphics[width=3.4in]{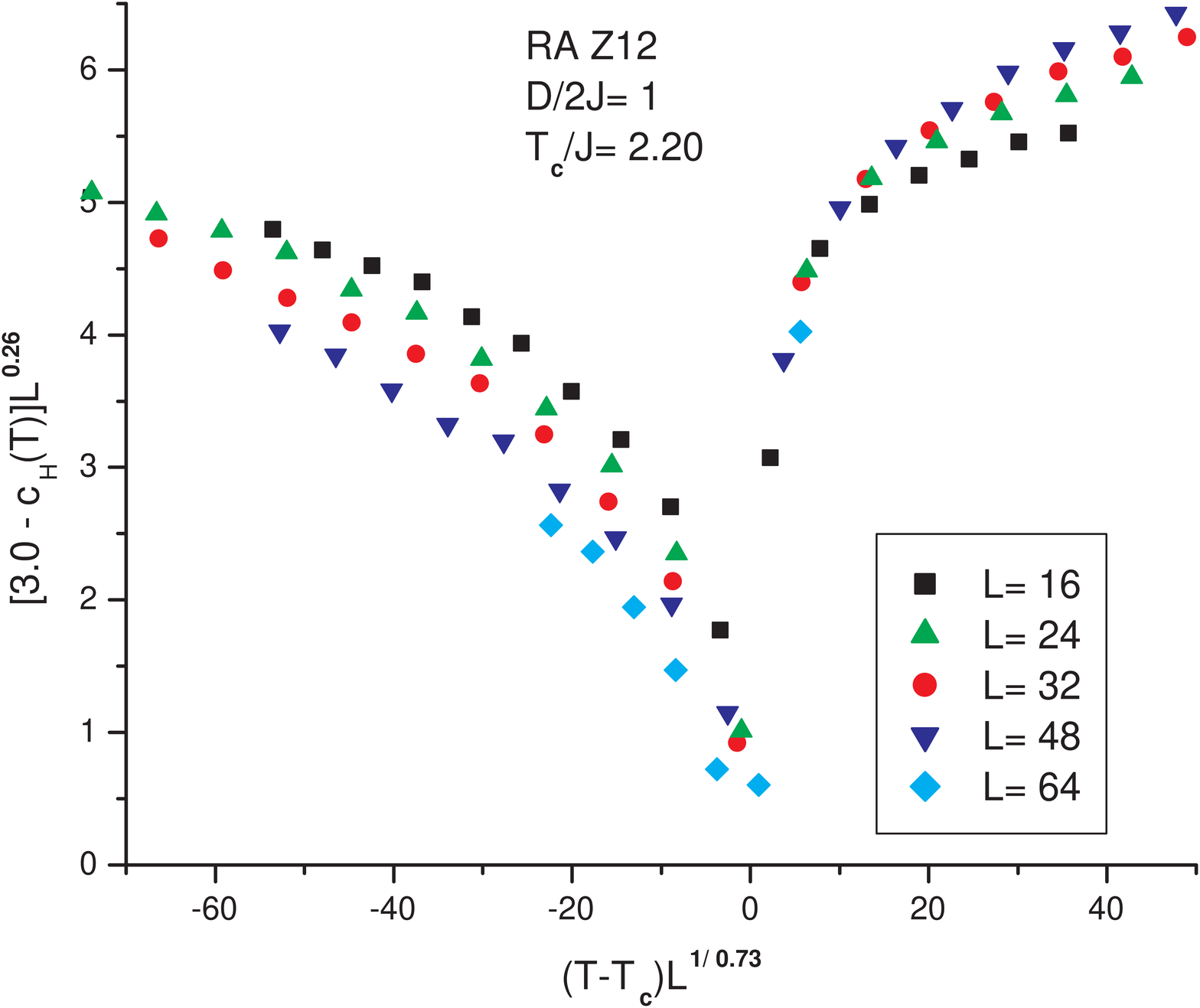}
\caption{\label{Fig.11}(color online) Finite-size scaling of the difference between
$c_H ( T_c )$= 3.00 and the configuration-averaged $c_H ( L, T )$ vs. temperature
near $T_c$ for $L \times L \times L$ lattices with $D / 2 J = 1$.}
\end{figure}

\begin{figure}
\includegraphics[width=3.4in]{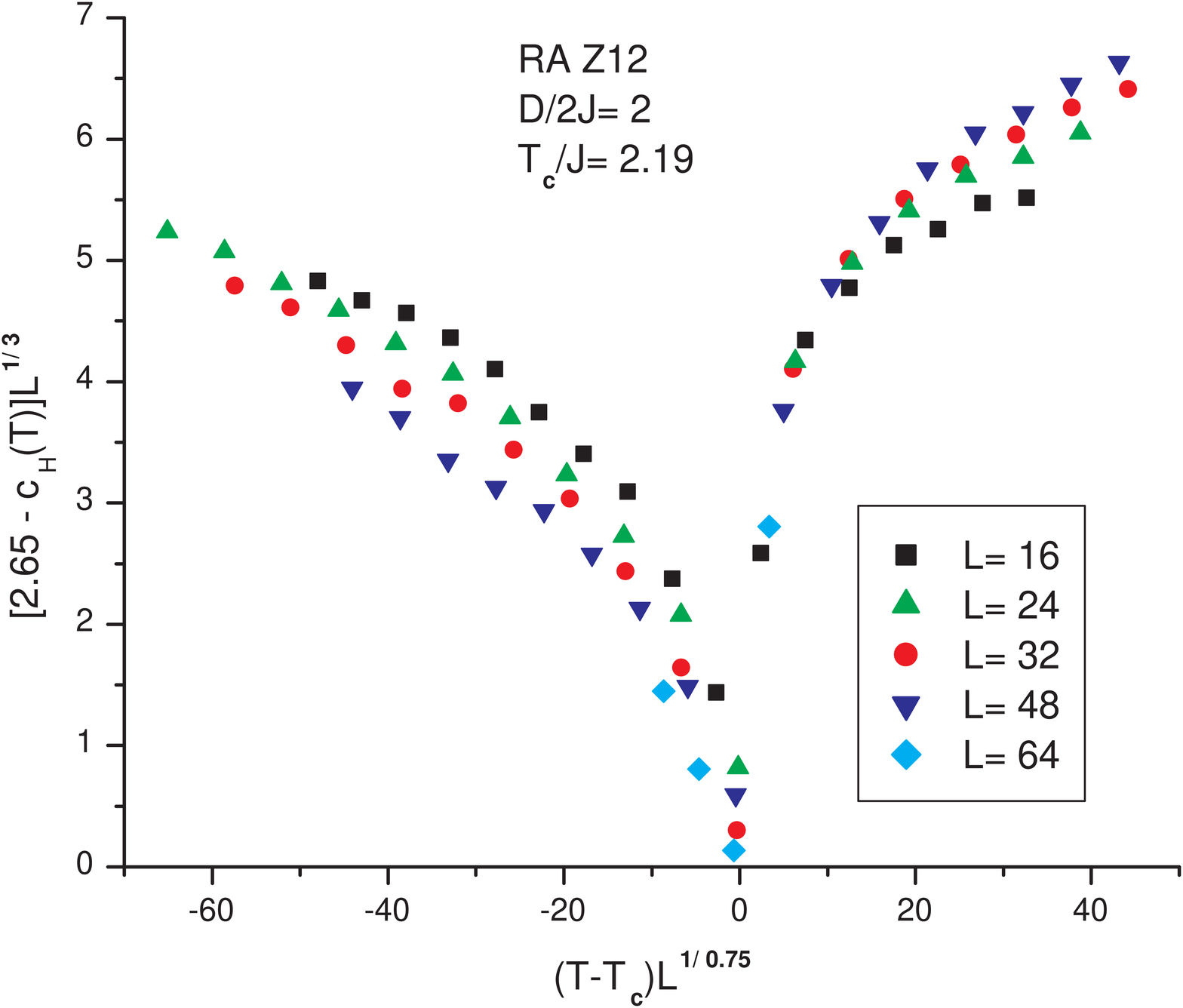}
\caption{\label{Fig.12}(color online) Finite-size scaling of the difference between
$c_H ( T_c )$= 2.65 and the configuration-averaged $c_H ( L, T )$ vs. temperature
near $T_c$ for $L \times L \times L$ lattices with $D / 2 J = 2$.}
\end{figure}

\begin{figure}
\includegraphics[width=3.4in]{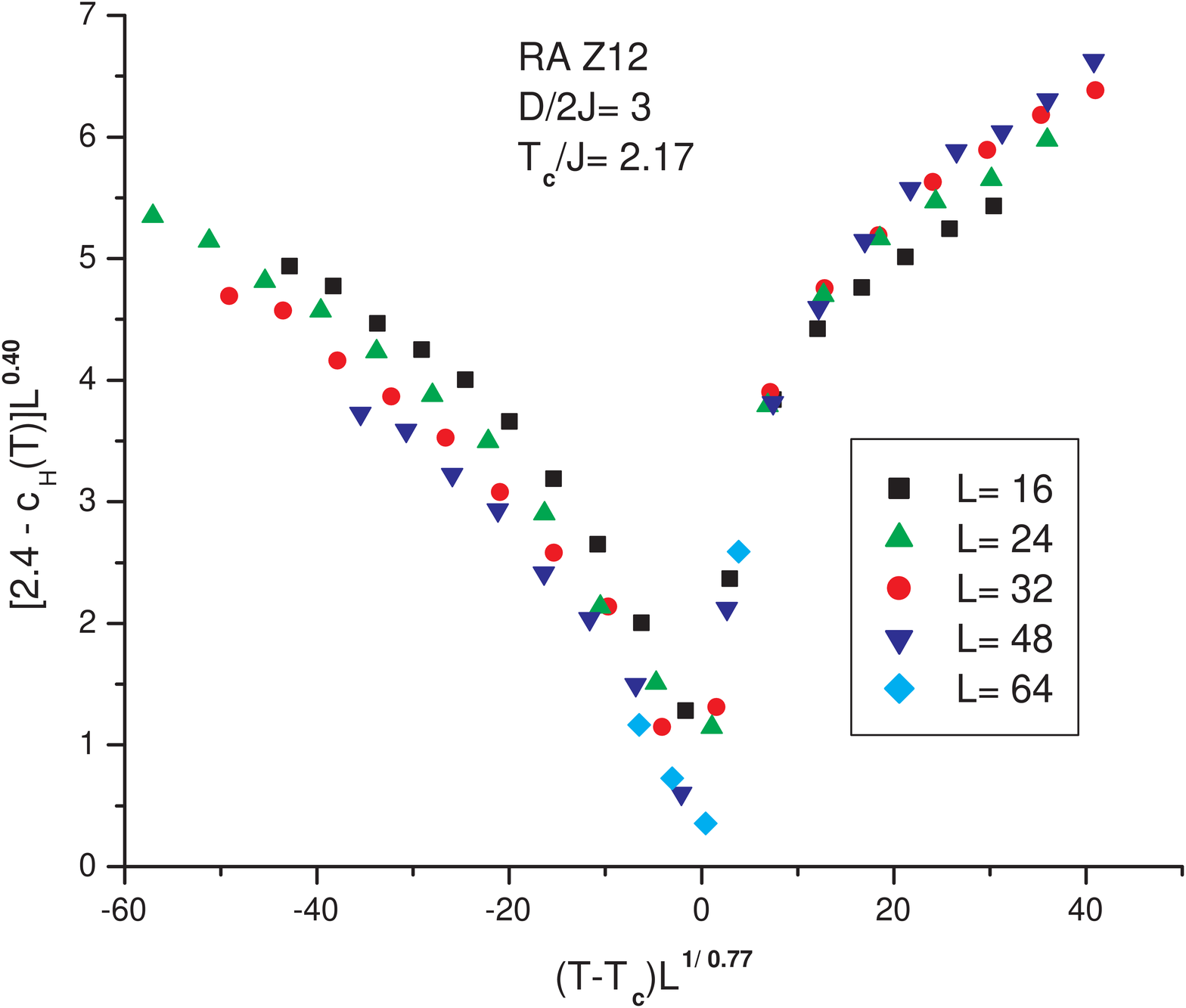}
\caption{\label{Fig.13}(color online) Finite-size scaling of the difference between
$c_H ( T_c )$= 2.40 and the configuration-averaged $c_H ( L, T )$ vs. temperature
near $T_c$ for $L \times L \times L$ lattices with $D / 2 J = 3$.}
\end{figure}

\begin{figure}
\includegraphics[width=3.4in]{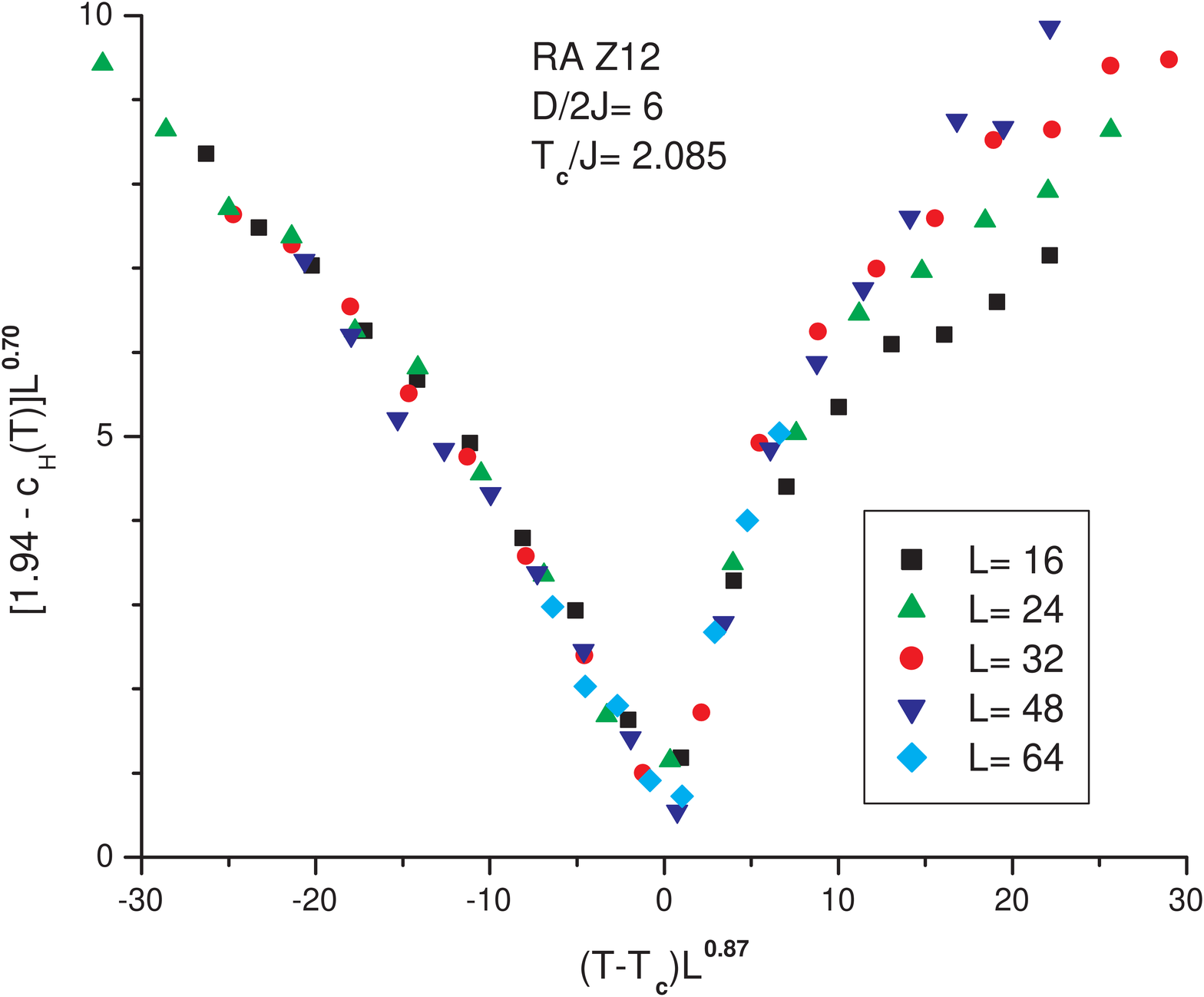}
\caption{\label{Fig.14}(color online) Finite-size scaling of the difference between
$c_H ( T_c )$= 1.94 and the configuration-averaged $c_H ( L, T )$ vs. temperature
near $T_c$ for $L \times L \times L$ lattices with $D / 2 J = 6$.}
\end{figure}

\begin{figure}
\includegraphics[width=3.4in]{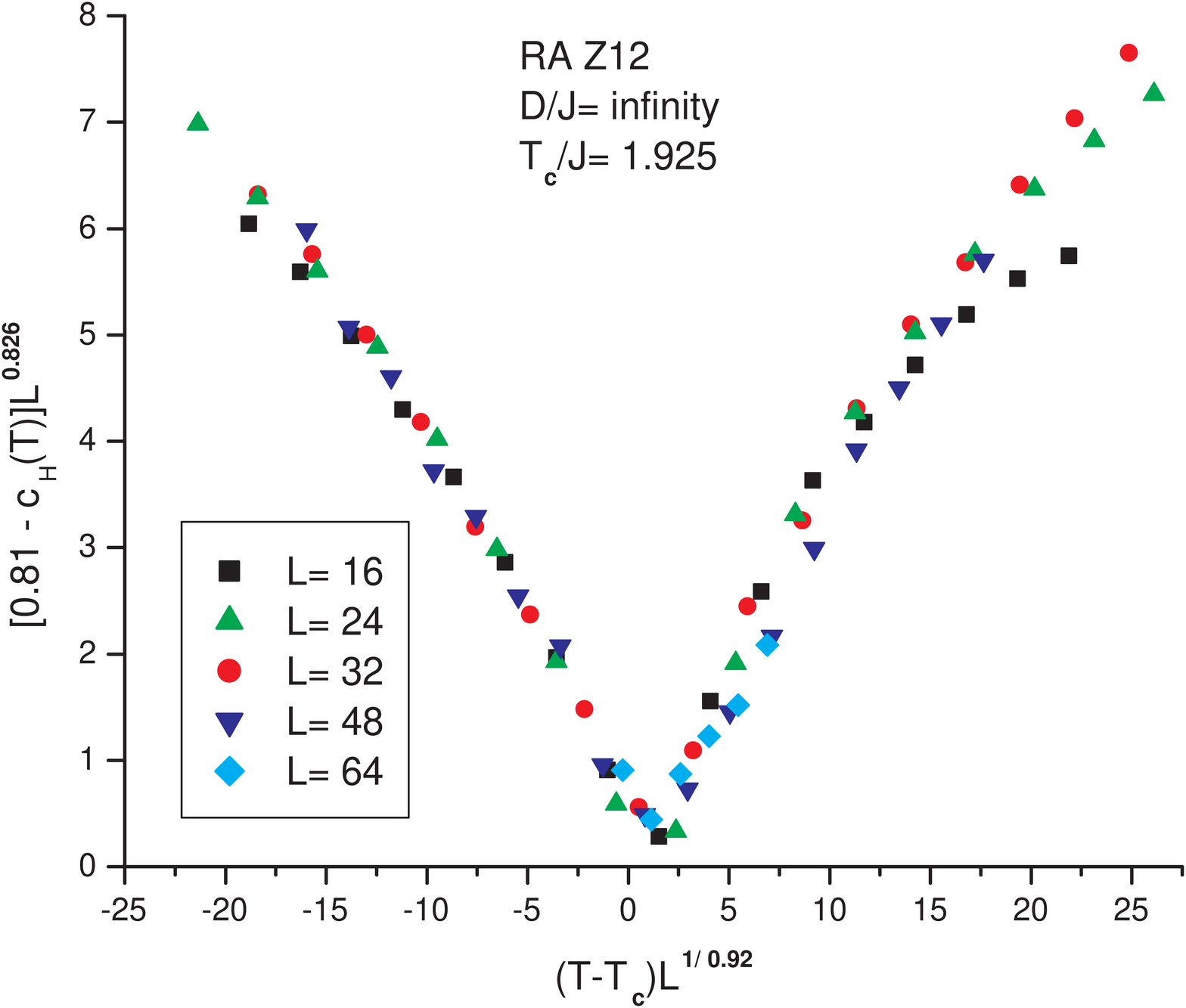}
\caption{\label{Fig.15}(color online) Finite-size scaling of the difference between
$c_H ( T_c )$= 0.81 and the configuration-averaged $c_H ( L, T )$ vs. temperature
near $T_c$ for $L \times L \times L$ lattices with $D = \infty$.}
\end{figure}

Fig.~11 shows the difference between an estimate of the specific heat at $T_c$
for an infinite system, $c_H ( T_c )$, and the calculated specific heat of a
finite system at temperature $T$, $c_H ( L, T )$ for $D / 2 J$ = 1.  The only
new adjustable fitting parameter here is $c_H ( T_c )$.  Figs.~12, 13, 14 and
15 show the corresponding plots for $D / 2 J$ = 2, 3, 6 and $\infty$,
respectively.  The values of $c_H ( T_c )$ decrease monotonically as $D / 2 J$
increases.  In all cases, the values of $c_H ( T_c )$ given in the figures are
estimated to be accurate to about 1\%. As we also saw for $|\vec{M}|$ and
$\chi_l$, the FSS data collapse is not good below $T_c$ for $D / 2 J \le
3$.  The results for $D = \infty$ are in very good agreement with the earlier
results\cite{Fis95} obtained with the $Z_6$ approximation.

\section{Discussion}

According to Imry and Ma\cite{IM75} and Pelcovits, Pytte and
Rudnick\cite{PPR78}, for small $D / J$ this model should appear
ferromagnetic when $L$ is smaller than the "Imry-Ma length", which is
determined by balancing the domain wall energy against the energy of
random pinning.  If this length exists, when $L$ is larger than the
Imry-Ma length, the system will break up into domains, without
long-range order.  We have argued here, however, that in the presence
of random pinning one should not believe that the domain wall energy
scales as $L^{d - 2}$.  One can try to patch up this picture by
assuming that the domain wall energy scales as $L^{d - \sigma_{dw}}$,
with $3/2 < \sigma_{dw} < 2$.  If this were the case, then it would
still be possible for $d = 3$ to find a length scale where the domain
wall energy balanced the random pinning energy.  Then it would continue
to be true in $d = 3$ that the system would break up into Imry-Ma-like
domains when $L$ became very large.

What we see in our FSS plots for $D / 2 J \le 3$, however, is that
when $T < T_c$ the leading correction to finite-size scaling increases
the magnetization as $L$ increases.  Therefore this model appears to be
stable against domain formation for $D / 2 J \le 3$, at least for some
range of $T$ below $T_c$.  The natural interpretation of this result is
that $\sigma_{dw}$ must be less than 3/2 in $d = 3$ for the $n = 2$ case.

Fig.~12, the FSS magnetization plot for $D / 2 J$ = 3, shows that as $L$
increases the data for $T < T_c$ seem to be converging to a scaling function
which is independent of $L$.  For the data in Fig.~15 for $D / 2 J$ = 6,
the data appear to be in this $L$-independent limit.  If we were able to do
the Monte Carlo calculations at substantially larger values of $L$, we would
expect to see the same type of convergence for $D / 2 J$ = 1 and 2.

If we had data at such large values of $L$, so that the magnetization
scaling function had converged to an $L$-independent limit, then our
estimates of the critical exponents would be expected to shift
somewhat.  Therefore, it is likely that $\eta^*$, the true value of $\eta$
in the range of $D / 2 J$ from 1 to 6, is actually independent of $D / J$.

The reader must also remember that the ferromagnetic phase is allowed
to be reentrant.  Therefore, we do not claim that the ferromagnetic
behavior which we see below $T_c$ must be stable down to $T$ = 0 over
the entire range of $D / J$.  Also, we do not claim stable ferromagnetism
for very large values of $D / J$.  It must be stated, however, that this
only applies to the simple cubic lattice with nearest neighbor interactions.
We expect that it would be possible to stabilize a ferromagnetic phase at
$D = \infty$ by adding further neighbor finite-range exchange interactions.

We point out that our earlier claim\cite{Fis95} of infinite magnetic
susceptibility without ferromagnetism when $D = \infty$ was based on
results at $T = 0$.\cite{Fis91}  Since the magnetization of finite simple
cubic lattices with $D = \infty$ seems to be a monotonically decreasing
function of $T$,\cite{Fis95} however, we consider the existence of true
ferromagnetism on this lattice to be unlikely at any $T$ for $D = \infty$.

The author sees no reason to believe that the exponent $\sigma_{dw}$
should be independent of $n$ for $d = 3$.  Thus, while we claim the
existence of a ferromagnetic phase for $n = 2$, we are not making any
claim here about the behavior for $n = 3$.  We do expect that
$\sigma_{dw}$ must converge to 2 in the limit $n \to \infty$, in
agreement with the result of Larkin.\cite{Lar70}  The reason for this is
that for $n \to \infty$ the "elastic membrane" approximation becomes valid.

Clearly, it would be desirable to obtain a direct estimate of $\sigma_{dw}$,
by, for example, calculating the change in energy of a sample between
periodic and antiperiodic boundary conditions along one direction.  Since
the energies involved are subextensive and the domain wall energy goes to
zero at $T_c$, it is difficult to do such a calculation.

\section{Summary}

In this work we have presented Monte Carlo results for the $d = 3$ $XY$
random anisotropy model, Eqn.~(2), for several values of the anisotropy
strength $D / J$.  By studying the finite-size scaling behavior of $L
\times L \times L$ simple cubic lattices over the range $16 \le L \le 64$,
we find that, for values of $D / J$ which are not very large, there
appears to be a finite-temperature critical point at which the model
undergoes a transition into a ferromagnetic phase.  For this lattice at
very large $D / J$, the transition appears to be into a phase with QLRO,
but no true magnetization.

\begin{acknowledgments}
The author thanks the Physics Department of Princeton University for providing
use of the computers on which the data were obtained.

\end{acknowledgments}



\end{document}